# New Results for Diffusion in Lorentz Lattice Gas Cellular Automata

E. G. D. Cohen
and
F. Wang
The Rockefeller University
New York, NY 10021

**Abstract**

New calculations to over ten million time steps have revealed a more complex diffusive behavior than previously reported, of a point particle on a square and triangular lattice randomly occupied by mirror or rotator scatterers. For the square lattice fully occupied by mirrors where extended closed particle orbits occur, anomalous diffusion was still found. However, for a not fully occupied lattice the super diffusion, first noticed by Owczarek and Prellberg for a particular concentration, obtains for all concentrations. For the square lattice occupied by rotators and the triangular lattice occupied by mirrors or rotators, an absence of diffusion (trapping) was found for all concentrations, except on critical lines, where anomalous diffusion (extended closed orbits) occurs and hyperscaling holds for all closed orbits with *universal* exponents $d_f = \frac{7}{4}$ and $\tau = \frac{15}{7}$. Only one point on these critical lines can be related to a corresponding percolation problem. The questions arise therefore whether the other critical points can be mapped onto a new percolation-like problem, and of the dynamical significance of hyperscaling.

# 1 Introduction

In a number of previous papers the diffusive behavior of Lorentz Lattice Gas Cellular Automata (LLGCA) has been studied[1−6]. Here a point particle moves in (constant) discrete time steps on a discrete lattice from site to site, a number of which is occupied randomly by stationary scatterers which scatter the particle according to strictly deterministic rules. The nature of the diffusive process of the particle through the scatterers has been the object of these investigations. In this paper we will confine ourselves to the motion of a particle on a square or triangular lattice occupied by fixed scatterers which remain unchanged during the diffusion. We will consider two different scatterer models: the mirror model and the rotator model. In the mirror (rotator) model a particle is scattered upon collision with a mirror (rotator) to the right or the left, depending on whether the mirror (rotator) is a right or a left mirror (rotator), respectively (cf.fig.1). The fraction of right (left) scatterers (mirrors or rotators, respectively) on the lattice will be denoted by $C_R$ ($C_L$), so that $C_R + C_L = C$ is the total fraction of lattice sites occupied by scatterers, i.e., the concentration $C$ of the scatterers. We also choose the time step = the lattice distance = speed = 1. For a given random placement of the mirrors (rotators), a particle will describe – from a given initial position – a random-like walk through the lattice (see

fig.1). The diffusive behavior of the particle will be obtained by averaging over all possible random configurations of the mirrors (rotators). It can be characterized by a number of quantities, of which we will only consider the mean square displacement $\Delta(t)$ and the radial distribution function $\hat{P}(r,t)$, defined by:

$$\Delta(t) = < r^2(t) > \tag{1}$$

$$\hat{P}(r,t) = \sum_{\theta=0}^{2\pi} rP(\vec{r},t) \tag{2}$$

respectively. Here $\vec{r} = (r,\theta)$ is the position of the particle in polar coordinates with respect to the origin, so that $r = |\vec{r}|$ is the distance of the particle from the origin. The average in eq. (1) is over all random configurations of the scatterers at fixed $C_R$ and $C_L$ and the sum in eq.(2) over all possible angles consistent with the lattice. $P(\vec{r},t)$ is the probability to find the particle at the position $\vec{r}$ at time $t$, given that it was at the origin at time $t = 0$, so that $\hat{P}(r,t)$ is the probability to find the particle a distance $r$ away from the origin at time $t$.

From eq. (1), we can define a time dependent diffusion coefficient $D(t)$, by:

$$D(t) = \frac{\Delta(t)}{4t} \tag{3}$$

so that a diffusion coefficient $D$ exists if in

$$D = \lim_{t\to\infty} D(t) \tag{4}$$

the right hand side (r.h.s.) exists.

When the diffusion is normal (we will call it class I), $P(\vec{r},t)$ will be a Gaussian given by:

$$P_G(\vec{r},t) = \frac{1}{2\sqrt{\pi Dt}}\exp(-\frac{r^2}{4Dt}) \tag{5}$$

which implies $\Delta(t) \sim t$, but not vice versa (cf. table I).

## 2 The Mirror Model

For the mirror model – which was introduced by Ruijgrok and one of us[1] – it was found that, although $D$ existed, $P(\vec{r},t)$ was not Gaussian. This was called anomalous (or class II) diffusion[4] (cf.figs.2a-b). This non-Gaussian behavior is due to the presence of closed (periodic) orbits, especially near the origin. This is in contrast to normal (Gaussian) diffusion where there are no closed orbits, but only open trajectories of the particles. The time evolution of $\hat{P}$ is indicated in fig.2b; it is dominated by closed orbits and zig-zag motions (cf.fig.2c). Since the time at which orbits close can be arbitrarily large, (infinite) extended closed orbits can occur.

The question arises whether all trajectories eventually close for all $C$ and $C_R/C_L$. For $C = 1$ this question was answered affirmatively by a theorem of Bunimovich and Troubetzkoy[7]: if $C = 1$ and $C_R > 0$; $C_L > 0$ then all trajectories are closed with probability 1. In this case ($C = 1$) one even knows how fast the trajectories close since it has been shown that the probability $P_o(t)$ to find an open orbit after time $t$ is given by[8]:

$$P_o(t) \sim t^{-\frac{1}{7}} \tag{6}$$

so that the number of open orbits decreases in time according to eq.(6) and all trajectories eventually close. The result (6) was obtained by noting a connection between the dynamical problem considered here and a bond percolation problem on the square lattice (see below).

For $C < 1$, only numerical evidence exists[9] that

$$P_o(t) \sim \frac{C_1}{\log t + C_2} \tag{7}$$

where $C_1$ and $C_2$ are constants. (7) implies that all trajectories still close, albeit much slower than (6), as illustrated in fig.3 (cf.ref.[9]).

Earlier work suggested anomalous diffusion for all $C$, which leads to a dynamical phase diagram of the mirror model on the square lattice like that in fig.4a. This phase diagram was based on calculations extending typically to 4,000 - 10,000 time steps. The method of calculation is described in refs. [2-5] and in the few calculations done for a larger number of time steps the class II behavior was consistent with the error bars of the data points (cf.fig.2a)[1].

However, in a recent paper by Owczarek and Prellberg[10] it was shown that for the particular case $C = \frac{2}{3}$ and $C_R = C_L = \frac{1}{3}$, the diffusion process was not anomalous but super diffusive, since $D(t)$ grew logarithmically with $t$. This result was based on very (many months) long calculations of up to a million time steps, giving results with very small error bars. This prompted us to extend all our calculations to longer times, in order to see how prevalent this super diffusive behavior was and to correct our phase diagram. In order to reduce the calculation times and in addition avoid the use of periodic boundary conditions, we employed a modification of the method developed by Ziff, Cummings and Stell[11]. Here a virtual lattice of 65536 $\times$ 65536 sites is divided into small blocks (256 $\times$ 256 lattice sites each) where scatterers are initially put randomly only on that block where the particle starts its motion, while no scatterers will be put on all the other blocks until the particle enters. This provides an enormous saving in memory and (along with other efficiencies) allows a reduction in computer time from months to days for obtaining results with very small error bars without any boundary effects. The details will be given in the next paper[12].

Figs.5a-b show our results for $D(t)$ for $C = 1$ as well as the decrease in the number

[1]A special case arises when $C_R$ (or $C_L$)=0 and $0 < C_L$ (or $C_R$)$< 1$, and (zig-zag) propagation occurs along the direction of the mirrors, while Gaussian (in fact, Boltzmann) diffusion takes place along the direction perpendicular to the mirrors.

$N_o(t)$ of open orbits for 10,000 particles for a number of values of $C_R/C_L$ up to a million time steps. The statistical errors were determined by doing the calculations in two steps: first an average was made over all 10,000 particles, with a different random configuration of the scatterers for each particle, then further averages were computed over typically three runs, involving three samples of 10,000 particles for each. The standard deviations of the mean are plotted as the error bars of the data in the figures. If the error bar does not appear, the error bar is inside the symbol. These results are consistent with class II behavior. Fig.6a shows $D(t)$ for a number of concentrations for $C < 1$ and $C_R = C_L$: contrary to what was found in fig.2a, $D(t)$ increases for sufficiently long times logarithmically in time for all $C < 1$, according to:

$$D(t) \sim A \log t \qquad (8)$$

where $A$ is essentially independent of $C_R$ for all $0 < C < 1$ (cf.fig.6b) (for $C$ very close to 0 or 1, we need to run for much longer times than feasible to obtain a constant slope). Fig.6c shows $D(t)$ for a number of concentrations $C_R/C_L$ for $C = 0.8$. The logarithmic increase (8) of $D(t)$ is clearly visible, but for $C_R \neq C_L$, the constant $A$ is $C_R$-dependent (cf.fig.6d). Thus, while the coefficient $A$ appears to be universal for $C_R = C_L$ it is not for $C_R \neq C_L$. As a result of the behavior of $D(t)$, the phase diagram for mirrors on a square lattice is not that of fig.4a, but that shown in fig.4b.

The $\hat{P}(r,t)$ for $C = 1$ and $C_L = C_R = 0.5$ is plotted in fig.7a, while that for $C = 0.6$ and $C_L = C_R = 0.3$ is plotted in fig.7b for a number of different times[2]. In both cases the curves increase in length and decrease in height with increasing time. Although the first case is that of anomalous diffusion (class II) and the second case of super-diffusion, the difference between the two curves is only significant around $r \approx 0$, where in the first case almost one order of magnitude more closed orbits occur than in the second case. For larger times, a difference is hardly noticeable and it is not clear whether this is due to the difference in concentration or in diffusive behavior.

We believe that the origin of the super diffusive behavior for $C < 1$ is two-fold:

1. the slow decay of the number of open trajectories, as given by eq.(7);

2. the possibility of large zig-zag motions (cf.fig.2c), because of the presence of unoccupied sites on the lattice.

# 3 Mirror Model and Percolation

For later discussions, we now sketch how the trajectories of the particle on the fully occupied square lattice (i.e., for $C = 1$) in the mirror model can be related to percolation clusters of a bond percolation problem on two sublattices of the original lattice[2,13]. In fig.8 the square lattice with two square sublattices is shown. We note that each lattice site of the original lattice is part of both sublattices[3]. Since only one mirror can be put at a lattice site, the mirror at this site can belong to only one of the two ambient sublattices. As a result, each sublattice is only half filled with mirrors, so that the probability $p$ that a bond in each sublattice is occupied by a mirror is $\frac{1}{2}$. This is just the critical value for bond percolation on a square lattice, so that the mirrors are at the bond percolation threshold on the two sublattices. Thus, if the size of the mirrors is chosen equal to the bond length of the sublattices, bond percolation clusters appear. We note that each closed orbit with period larger than 4 time steps is formed by reflection between the inner and outer perimeters of bond clusters on the sublattices (cf.fig.8). There is no ambiguity about "inner" or "outer", as the closed orbits do not cross themselves and the inner and outer perimeters refer to clusters on *different* sublattices. This mapping implies the following analogy. From percolation theory we know that the mean square gyration radius $< R_N^2 >_o$ of the perimeter of a percolation cluster with $N$ perimeter sites grows with $N$ as:

$$< R_N^2 >_o \sim N^{\frac{2}{d_f}} = N^{\frac{8}{7}} \tag{9}$$

[2]There is no obvious difference in behavior from fig.2b, except that we used periodic boundary conditions in fig.2b, rather than the procedure of ref.[11].

[3]The sublattices were plotted wrongly in fig.1 in ref.[2].

where the fractal dimension $d_f = \frac{7}{4}$. Furthermore the probability $P_o(N)$ to find an open percolation cluster with $N$ perimeter sites decreases with $N$ as:

$$P_o(N) \sim N^{2-\tau} = N^{-\frac{1}{7}} \tag{10}$$

so that the size distribution parameter $\tau = \frac{15}{7}$. We note that $d_f$ and $\tau$ satisfy a hyperscaling relation:

$$\tau - 1 = \frac{2}{d_f} \tag{11}$$

Similarly, for the trajectories of the particles on the lattice, one finds that the mean square displacement $\Delta_o(t)$ for particles on open trajectories at time $t$ increases as:

$$\Delta_o(t) \sim t^{\frac{8}{7}} \tag{12}$$

while the probability for an open trajectory at time $t$ is given by:

$$P_o(t) \sim t^{-\frac{1}{7}} \tag{6}$$

The analogy of (9), (12) and also (10), (6), respectively, is obvious and illustrated in fig.5b.

## 4 The Rotator Model

We now discuss our rotator model, which is a special case of a set of rotator models introduced by Gunn and Ortuño[14]. The scattering rules are illustrated in fig.1 and although similar to those of the mirror model, differ from those in their scattering results in half of the cases (cf. the rules in fig.1); they do not lead to time-reversible particle motions, as in the mirror model.

In earlier work, the rotator model seemed to behave for $C_R = C_L$ very similarly to the mirror model (cf.fig.9), while for $C_R \neq C_L$ a qualitative difference was noticed. Then, instead of showing a class II anomalous diffusive behavior as the mirror model did, a dynamical phase transition was observed, for sufficiently large $C_R$ or $C_L$, from anomalous to an absence of diffusion–to which we will refer as no-diffusion– where the mean square displacement became bounded ($\Delta(t)$ < constant) and all particles trapped, so that no extended closed orbits occurred. We called this class IV behavior[4]. This behavior is illustrated in fig.10a for $D(t)$ and in fig.10b for $\hat{P}(r,t)$. We note, that indeed for $t \geq 2^{10}$, the curve for $D(t)$ on $\log_{10}$-$\log_{10}$ scale has a slope of -1 and $\hat{P}$ appears stationary and does not seem to change anymore, as all particles are trapped by that time. This leads to a phase diagram as pictured in fig.11a, where the phase transition occurs across two (approximately determined) straight lines, anchored on the $C_R$ and $C_L$ axes at $C_R = C_L = 0.593$, the critical concentration for site percolation on the square lattice. The point $C_R = C_L = 0.5$ for $C = 1$, which can be mapped on the corresponding point for the mirror model, has class II behavior[2]. This result was consistent with another theorem proved by Bunimovich and Troubetzkoy[7] which stated that 1) there exists a critical concentration $C_{R_{cr}}$ or $C_{L_{cr}} \in (0.5, 1)$ (our 0.593), such that, for $C_R > C_{R_{cr}}$ or $C_L > C_{L_{cr}}$, all trajectories are periodic with probability 1; 2) for $C = 1$ all trajectories are periodic with probability 1. The proof is based on the observation that if an infinite percolation cluster of, say, right rotators exists, the particle inside the cluster is everywhere surrounded by a contour consisting of only one kind of rotators and consequently is trapped. The regions covered by this theorem are indicated in fig.11a.

Recently we have extended all calculations for this model also to a million and sometimes close to more than ten million time steps and again found very different results as were obtained before for typically 4,000 - 10,000 time steps.

In fig.12 $D(t)$ is plotted for $C = 1.0$ and 0.8 for $C_R = C_L$ as well as for a particular value $C_R \neq C_L$ for $C = 0.9$. While for $C_L = C_R$ and $C = 0.8$, $D(t) \to 0$, i.e., class IV behavior or trapping is observed, for the particular concentrations $C_R = 0.423$, $C_L = 0.477$ for $C = 0.9$, $D(t)$ seems to approach a constant $\neq 0$, i.e., class II behavior or extended closed orbits occur, like for $C_L = C_R = 0.5$ when $C = 1$. The possible occurrence of extended closed orbits for special values of $C_R$ and $C_L$ is confirmed by further study. Fig.13a shows the number of open orbits $N_o(t)$ for $C = 1$ for a variety of $C_R$ and $C_L$. Only for $C_R = C_L = 0.5$ do extended trajectories occur and a corresponding slow decay of $N_o(t)$ occurs, while for $C_R \neq C_L$, the precipitous decay of $N_o(t)$ clearly indicates the sudden trapping of particles when $t$ approaches a critical value. Similarly figs.13b-c show class II behavior only for the particular values $C_R = 0.455$, $C_L = 0.495$ and $C_R = 0.46$, $C_L = 0.39$ for $C = 0.95$ and $C = 0.85$, respectively. The number of closed orbits $N_c(t)$ at these critical concentrations of $C_R$ and $C_L$ is a minimum when compared to those at all other concentrations of $C_R$ and $C_L$ at a given value of $C$. For all these cases, the decay of $N_o(t)$ with time is again given by eq.(6), confirming the suspected class II behavior deduced from the behavior of $D(t)$. Fig.13d shows the similarity of the decay of $N_o(t)$ for a number of concentrations and suggests a dynamical phase diagram as given in fig.11b, instead of that in fig.11a. The existence of critical points for $C < 1$ was noticed before by Ortuño, Ruiz and Gunn[15], but the location and properties of their points are quite different from ours.

The present situation can be summarized as follows.

1. There appear to be two symmetric critical lines – when $C_R/C_L$ is critical, so is $C_L/C_R$ – in the phase diagram, where extended closed orbits of a particle can occur.

Outside these critical lines there is trapping everywhere.

2. Only for $C = 1$ is there a connection of the rotator model with percolation through a mapping on a bond percolation problem[14].

In fact, at $C = 1$ a given closed orbit can always be mapped into the perimeter of a bond cluster as follows: we replace each rotator along the trajectory by a bond which looks like a mirror with a length of $\sqrt{2}$ lattice distances and with an orientation such that the bond does not cross the particle trajectory (cf.fig.8). Like in the mirror model, the particle trajectory bounces back and forth between the inner and outer perimeters of bond clusters. Like for the mirror model, there is no ambiguity about "inner" and "outer" perimeters since the closed orbit does not cross itself and the bonds do not cross the closed orbits either and the inner and outer perimeters of bond clusters reside on different sublattices, respectively (cf.fig.8). However, unlike for the mirrors, the inner perimeter of a bond cluster always corresponds to one kind of rotators (R or L) while the outer perimeter of bond cluster always corresponds to the other kind of rotators (L or R).

We will now use this to argue that for $C = 1$ percolation occurs only for $C_L = C_R = 0.5$, unlike for the mirrors. First we note that the number of (the same kind of) rotators corresponding to the perimeter of the inner bond cluster is always smaller than that of the corresponding outer bond cluster. Because the particle trajectory has a one to one relation to the inner bond cluster, it follows that for large closed orbits the probability for a bond of the inner perimeter of bond cluster will depend on the smaller of the two concentrations $C_R$ and $C_L$ at $C = 1$. Therefore, since $C_L = C_R = 0.5$ just corresponds to the bond percolation cluster threshold, where the probability $p = p_c = 0.5$, so that there will exist extended closed orbits, leading to a diffusive behavior of class II. However, for $C_L \neq C_R$, the perimeter of the inner bond cluster corresponds to a $p < p_c = 0.5$, so that there is then no inner bond percolation cluster and there are no extended closed orbits, i.e., all particles are trapped and the diffusive behavior is that of class IV.

For $C < 1$, no such mapping seems to exist, since no percolation clusters can be formed because of the empty sites on the lattice. Nevertheless, we determined *independently* that $\tau = \frac{15}{7}$ by $P_o(t) \sim t^{2-\tau} = t^{-\frac{1}{7}}$ (cf.figs.13a-d) as well as $d_f = \frac{7}{4}$ from $P_o(t)\Delta_o(t)/t \sim t^{-\frac{1}{7}}t^{\frac{2}{d_f}}/t = t^{-\frac{1}{7}}t^{\frac{8}{7}}/t \sim$ constant (cf.fig.14a) leading, within the experimental errors, to the validity of the hyperscaling relation eq.(11) along the two symmetric critical lines.

3. In spite of what one would expect physically, for $C < 1$, there appears to be no obvious influence of the site percolation transition and the formation of an infinite cluster of right (or left) rotators, on which the proof of the above mentioned Bunimovich - Troubetzkoy theorem is based. The Bunimovich - Troubetzkoy theorem seems therefore only sufficient, not necessary for trapping of the particle (class IV behavior) to occur.

## 5 The Mirror and Rotator Model on the Triangular Lattice

A piece of a triangular lattice is shown in fig.15a. The motion of the particle takes place on a triangular lattice if the scattering of the particle occurs over the largest possible angle $2\pi/3$, but on a honeycomb (sub) lattice, if the scattering occurs over the smallest angle $\pi/3$[3]. The latter case will be discussed in the following paper[12]. Right and left mirrors as well as rotators can be defined on the triangular lattice as illustrated in fig.15b.

For relatively short times ($t \leq 10,000$), we found for both (mirror and rotator) models the dynamical phase diagram of fig.16a, similar to that for rotators on the square lattice (cf.fig.11a). The only difference is that the critical concentration for site percolation on the triangular lattice is $C_{R_{cr}} = C_{L_{cr}} = 0.5$ instead of 0.593. A similar theorem as for the rotator model on the square lattice has been proved for the mirror and rotator models on the triangular lattice by Bunimovich and Troubetzkoy[7]. The regions covered by this theorem are indicated in fig.16a.

The recently carried out extended calculations up to a million or more time steps again showed very different results.

In fig.17a, $D(t)$ is plotted for a number of values of $C_L$ and $C_R$ showing that, for $C_L = C_R$, $D(t)$ approaches a constant, while for $C_L \neq C_R$ it decays with a slope of $-1$, indicating, with eq.(3), that the mean square displacement is bounded. There is again the occurrence of a critical concentration at $C_L = C_R$ for $C = 0.8$ as illustrated in fig.17b, where again the precipitous decay of $N_o(t)$ for $C_L \neq C_R$ clearly indicates the sudden trapping of particles for a finite $t$. The slow power-law decay of $N_o(t)$ according to eq.(6) for $C_R = C_L = 0.4$ is due to the occurrence of (infinitely) extended closed orbits and holds for all $C_L = C_R$. Fig.17c shows this behavior for a number of different $C$ for $C_R = C_L$ and fig.16b gives the new phase diagram, which is – for the triangular lattice – identical for the mirror and the rotator models.

This same behavior of mirror and rotator model can be understood by replacing all the right (left) rotators on a particle trajectory by either right (left) mirrors or left (right) mirrors, while not changing the empty sites on the particle trajectory. Thus in fig.18, we only need to replace all right mirrors by right (left) rotators and all left mirrors by left (right) rotators, to obtain the same trajectory with the same (opposite) direction of particle motion, as in fig.18, respectively. So, the mapping will be either $C_L^{rotator} = C_L^{mirror}$, $C_R^{rotator} = C_R^{mirror}$ or $C_L^{rotator} = C_R^{mirror}$, $C_R^{rotator} = C_L^{mirror}$. There is no difference between these two cases since the diffusive behavior of the models is invariant for an interchange of $C_L$ and $C_R$.

We notice in fig.16b that there is now only one critical line $C_R = C_L$, which exhibits class II behavior. Like for the square lattice we determined independently the exponents $\tau = \frac{15}{7}$ and $d_f = \frac{7}{4}$ (cf.figs.14b and 17b-c, respectively) leading, within the experimental errors, to a verification of the hyperscaling relation eq.(11) along the entire critical line. For all other concentrations class IV behavior obtains. Only for $C = 1$ for both (mirror and rotator) models is there a mapping to a site percolation problem on the same lattice obtained by connecting the same kind of nearest neighbor scatterers (mirrors or rotators) (cf.fig.18).

The present situation can be summarized as follows.

1. There appears to be one critical line in the phase diagram where (infinite) extended closed orbits of a particle occur and hyperscaling relation eq.(11) holds. Outside this critical line there is everywhere trapping.

2. Only for $C = 1$ for both (mirror and rotator) models is there a connection with site percolation on the *same* lattice as that of the particle trajectory (fig.18) (cf. in section **4**, the mapping of the particle trajectories for the rotator model on the square lattice onto bond clusters on dual lattices).

3. For $C < 1$, there appears no noticeable influence of the percolation transition, on which the above mentioned proof of Bunimovich-Troubetzkoy theorem is based.

Thus, the Bunimovich-Troubetzkoy theorem seems also in this case only a sufficient, not a necessary condition, for class IV behavior to occur.

# 6 Outlook

We summarize our new results in table II.

We close with the following questions and remarks.

1. The most striking result of these investigations seems to be the existence for both lattices of critical lines which are extentions of percolation transitions and appear to have universal percolation perimeter critical exponents satisfying the same hyperscaling relation eq.(11) as found for the percolation problem. On these critical lines extended closed orbits occur, but what determines for a given concentration $C$ the critical concentrations for these extended orbits to occur, is unknown to us.

2. Are there for a not fully occupied lattice ($C < 1$) generalized percolation cluster-like structures to which the extended closed orbits are geometrically related, like the percolation clusters on a fully occupied lattice ($C = 1$), i.e., does our dynamical model suggest something new – a possible generalization – for the percolation problem?

3. What is the dynamical meaning of hyperscaling relations for non-percolation related trajectories?

4. Can one understand that in all cases –, i.e., on both the square and the triangular lattice and for both mirror and rotator models (except for the mirror model on the square lattice for $C_R$ or $C_L = 0$) – the trajectory of a moving particle eventually always closes? Furthermore, how can one see that in most cases the scatterers will lead to a quick trapping of the particle, but that in some – rather special – cases closed orbits of any extent can occur so that the closing of orbits proceeds power-law or logarithmically slow. We note that the Bunimovich-Troubetzkoy theorems proved so far do not distinguish between class II and class IV closing of orbits.

5. In all cases considered here, all trajectories close eventually. This does not imply that no-diffusion takes place in any form; for anomalous (class II) diffusion, one can still define a finite diffusion coefficient by the relation (4). In fact, a whole range of possible diffusive behaviors occurs ranging from super-diffusion to no-diffusion. The closing of all trajectories does therefore not by itself say anything about the way the particles diffuse – as measured by their mean square displacement – through the scatterers.

6. The previously obtained results for the flipping mirror and rotator models[5] do not seem to be affected by the present extension of the time scale of our calculations. They remain therefore unchanged.

## Acknowledgement

The authors are grateful to Robert Ziff for many very helpful discussions. Part of this work was supported by the Department of Energy under grant No. DE-FG02-88ER13847.

**Figure Captions**

Fig.1 Typical mirror (a) and rotator (b) configurations, particle trajectories and scattering rules for the mirror and rotator models, respectively, on the square lattice.

Fig.2 (a) Old calculation of the diffusion coefficient $D$ as a function of the time $t$ on a $\log_{10}$-$\log_{10}$ scale for the fixed mirror model on the square lattice for $C_L = C_R = 0.25$[2]; (b) corresponding radial density distribution functions $\hat{P}(r,t)$ as a function of distance $r$ from the origin at time steps $t = 2^{10}$ ($\diamond$), $t = 2^{12}$ ($+$) and $t = 2^{14}$ ($\square$), respectively. The stationary peak near the origin is due to closed orbits; (c) a possible closed orbit (A) and zig-zag motion (B) for the mirror model on the square lattice for $C < 1$.

Fig.3 Inverse probability for open orbits as a function of $\log_2 t$ for the mirror model on the square lattice for $C_L = C_R = 0.4$.

Fig.4 (a) Old phase diagram for the mirror model on the square lattice[2]; (b) new phase diagram.

Fig.5 (a) Diffusion coefficient $D$ as a function of $\log_{10} t$ for the mirror model on the square lattice for $C = 1$ at $C_L = C_R = 0.5$ ($\diamond$), $C_L = 0.55$, $C_R = 0.45$ ($+$) and $C_L = 0.75$, $C_R = 0.25$ ($\square$) (the first two overlap); (b) number of closed orbits as a function of $t$ on a $\log_2$-$\log_2$ scale for the mirror model on the square lattice for $C = 1$ at $C_L = 0.6$, $C_R = 0.4$ ($\diamond$) and $C_L = 0.5$, $C_R = 0.5$ ($+$). The lines through the points are drawn to guide the eye, virtually coincide. The slope for both curves $\approx -1/7$.

Fig.6 (a) Diffusion coefficient $D$ as a function of $\log_{10}t$ for the mirror model on the square lattice for $C_L = C_R = 0.4$ ($\diamond$), $C_L = C_R = 0.35$ (+) and $C_L = C_R = 0.3$ ($\square$); (b) slope of $D/\log_{10}t$ as a function of the concentration of mirrors $C$ for $C_L = C_R$; (c) as in (a) for $C_L = C_R = 0.4$ ($\diamond$), $C_L = 0.5$, $C_R = 0.3$ (+), $C_L = 0.6$, $C_R = 0.2$ ($\square$), $C_L = 0.7$, $C_R = 0.1$ ($\times$) and $C_L = 0.75$, $C_R = 0.05$ ($\triangle$); (d) slope of $D/\log_{10}t$ as a function of concentration of left mirrors $C_L$ for $C = 0.8$ and $C_L \neq C_R$. The lines through the points are drawn to guide the eye.

Fig. 7 Radial distribution functions $\hat{P}(r,t)$ for the mirror model on the square lattice as a function of the distance $r$ from the origin at time steps $t = 2^{12}$ ($\diamond$), $t = 2^{14}$ (+) and $t = 2^{16}$ ($\square$), respectively, (a) for $C = 1$ at $C_L = C_R = 0.5$ (peak value of about 0.27 is not shown in the figure); (b) for $C = 0.6$ at $C_L = C_R = 0.3$.

Fig 8. A typical particle trajectory (thin solid line with arrows) on the square lattice (thin solid lines) relates to the perimeters of bond clusters each on one of two sublattices, respectively (dashed and dotted lines, respectively). The thick lines are the perimeters of the bond clusters (also the mirrors for the mirror model), while "L" and "R" refer to rotators for the corresponding situation for the rotator model. The particle trajectory resides between the inner and outer perimeters of the two clusters.

Fig.9 Old calculation of diffusion coefficient $D$ as a function of $t$ on a $\log_{10}$-$\log_{10}$ scale for the rotator model on the square lattice for $C_L = C_R = 0.25$[4].

Fig.10 (a) Diffusion coefficient $D$ as a function of the time $t$ on a $\log_{10}$-$\log_{10}$ scale for the rotator model on the square lattice for $C_L = 0.6$, $C_R = 0.2$; (b) corresponding radial distribution functions $\hat{P}(r,t)$ as a function of distance $r$ from the origin at time steps $t = 2^7$ ($\diamond$), $t = 2^{10}$ ($+$) and $t = 2^{13}$ ($\square$), respectively.

Fig.11 (a) Old phase diagram for the rotator model on the square lattice[4]; the part with dashed lines corresponds to that covered by the theorem of Bunimovich and Troubetzkoy; (b) new phase diagram. The dash-dotted lines indicate the extrapolated behavior we expect for $C < 0.7$, the minimum value of $C$ we considered.

Fig.12 Diffusion coefficient $D$ as a function of $t$ on a $\log_{10}$-$\log_{10}$ scale for the rotator model on the square lattice for the critical concentrations for $C = 0.9$ at $\mathbf{C_L = 0.477}$, $\mathbf{C_R = 0.423}$ ($\diamond$), $C = 1$ at $\mathbf{C_L = C_R = 0.5}$ ($+$) and $C = 0.8$ at $C_L = C_R = 0.4$ ($\square$); critical concentrations are printed in bold face.

Fig.13 Number of open orbits out of 10,000 trajectories as a function of $t$ on a $\log_2$-$\log_2$ scale for the rotator model on the square lattice. (a) $C = 1$ at $\mathbf{C_L = C_R = 0.5}$ ($\diamond$), $C_L = 0.51$, $C_R = 0.49$ ($+$), $C_L = 0.52$, $C_R = 0.48$ ($\square$), $C_L = 0.53$, $C_R = 0.47$ ($\times$) and $C_L = 0.54$, $C_R = 0.46$ ($\triangle$) (the number of closed orbits has a minimum at $C_L = C_R = 0.5$); (b) $C = 0.95$ at $C_L = C_R = 0.475$ ($\diamond$), $C_L = 0.485$, $C_R = 0.465$ ($+$), $\mathbf{C_L = 0.495, C_R = 0.455}$ ($\square$) and $C_L = 0.505$, $C_R = 0.445$ ($\times$) (the number of closed orbits has a minimum at $C_L = 0.495$, $C_R = 0.455$); (c) $C_L = C_R = 0.425$

($\diamond$), $C_L = 0.43$, $C_R = 0.42$ ($+$), $C_L = 0.44$, $C_R = 0.41$ ($\square$), $C_L = 0.45$, $C_R = 0.4$ ($\times$), $\mathbf{C_L = 0.46}$, $\mathbf{C_R = 0.39}$ ($\triangle$) and $C_L = 0.47$, $C_R = 0.38$ ($*$) (the number of closed orbits has a minimum at $C_L = 0.46, C_R = 0.39$); (d) same as in (c) for $\mathbf{C_L = C_R = 0.5}$ ($\diamond$), $\mathbf{C_L = 0.495, C_R = 0.455}$ ($+$), $\mathbf{C_L = 0.477, C_R = 0.423}$ ($\triangle$), $\mathbf{C_L = 0.46, C_R = 0.39}$ ($\square$) and $\mathbf{C_L = 0.44, C_R = 0.36}$ ($\times$). The lines through the points are drawn to guide the eye.

Fig.14 Contribution of diffusion coefficient from open orbits $P_o(t)\Delta_o(t)/t$ as a function of time $t$ on a $\log_2$-$\log_2$ scale for (a) $C = 0.85$ at $\mathbf{C_L = 0.46}$, $\mathbf{C_R = 0.39}$ ($\diamond$), $C_L = 0.45$, $C_R = 0.4$ ($+$) and $C_L = 0.47$, $C_R = 0.38$ ($\square$) for the rotator model on the square lattice; (b) $C = 0.8$ at $\mathbf{C_L = C_R = 0.4}$ ($\diamond$) and $C_L = 0.41$, $C_R = 0.39$ ($+$) for the rotator model on the triangular lattice. The approach to a horizontal line is apparent for the critical concentrations.

Fig.15 (a) A typical particle trajectory on the triangular lattice, three-sided dotted lines stand for right and left mirrors and "R" and "L" stand for right and left rotators, respectively; (b) the scattering rules for the mirror and rotator models on the triangular lattice.

Fig.16 (a) Old phase diagram for both mirror and rotator model on the triangular lattice, dashed lines as in fig.12(a); (b) new phase diagram.

Fig.17 (a) Diffusion coefficient $D$ as a function of $t$ on a $\log_{10}$-$\log_{10}$ scale for the ro-tator model on the triangular lattice for $C = 1$ at $\mathbf{C_L = C_R = 0.5}$ ($\triangle$), $C_L = 0.45$, $C_R = 0.55$ ($\times$), $C = 0.8$ at $\mathbf{C_L = C_R = 0.4}$ ($*$), $C = 0.79$ at $C_L = 0.4$, $C_R = 0.39$ ($\diamond$), $C = 0.65$ at $\mathbf{C_L = C_R = 0.325}$ ($+$) and $C = 0.5$ at $\mathbf{C_L = C_R = 0.25}$ ($\square$); (b) number of open orbits out of 10,000 trajectories as a function of $t$ on a $\log_2$-$\log_2$ scale for the rotator model on the triangular lattice for $C = 0.8$ at $\mathbf{C_L = C_R = 0.4}$ ($\diamond$), $C_L = 0.41$, $C_R = 0.39$ ($+$), $C_L = 0.43$, $C_R = 0.37$ ($\square$), $C_L = 0.44$, $C_R = 0.36$ ($\times$) and $C_L = 0.45$, $C_R = 0.35$ ($\triangle$) (the number of closed orbits has a minimum at $C_L = C_R = 0.4$); (c) same as in (b) for $\mathbf{C_L = C_R = 0.5}$ ($\diamond$), $\mathbf{C_L = C_R = 0.4}$ ($+$), $\mathbf{C_L = C_R = 0.3}$ ($\square$) and $\mathbf{C_L = C_R = 0.25}$ ($\times$). The lines through the points are drawn to guide the eye.

Fig.18 A typical particle trajectory (closed orbit) on the triangular lattice (thin and thick solid lines with arrows) relates to site clusters on the same triangular lattice as for the particle trajectory. The thick lines are the perimeters of the site clusters. The three-sided dotted lines represent the mirrors for the mirror model, "L" and "R" represent the rotators for the rotator model and the particle trajectory is between the inner and outer perimeters of the clusters. We can see that the mirror model and the rotator model are the same by replacing the right (left) mirrors by right (left) rotators or vice versa.

## Table Captions

Table I – Comparison of normal and anomalous diffusion

Table II – Comparison of different diffusive behavior

**Table I – Comparison of normal and anomalous diffusion**

| | Normal (class I) | Anomalous (class II) |
|---|---|---|
| $\Delta(t)$ | $\sim t$ | $\sim t$ |
| $P(\vec{r},t)$ | Gaussian | Non-Gaussian |

**Table II – Comparison of different diffusive behavior**

| Lattice | MIRROR | ROTATOR |
|---|---|---|
| Square | $0 < C < 1 \rightarrow$ Super-diffusion | $0 < C \leq 1 \rightarrow$ No-diffusion (class IV) except for 2 critical lines $\rightarrow$ Anomalous diffusion (class II) |
| | $C = 1 \rightarrow$ Anomalous diffusion (class II) | |
| | see fig.4b | see fig.11b |
| Triangular | $C_L \neq C_R \rightarrow$ No-diffusion (class IV) | |
| | $C_L = C_R \rightarrow$ one critical line $\rightarrow$ Anomalous diffusion (class II) | |
| | see fig.16b | |

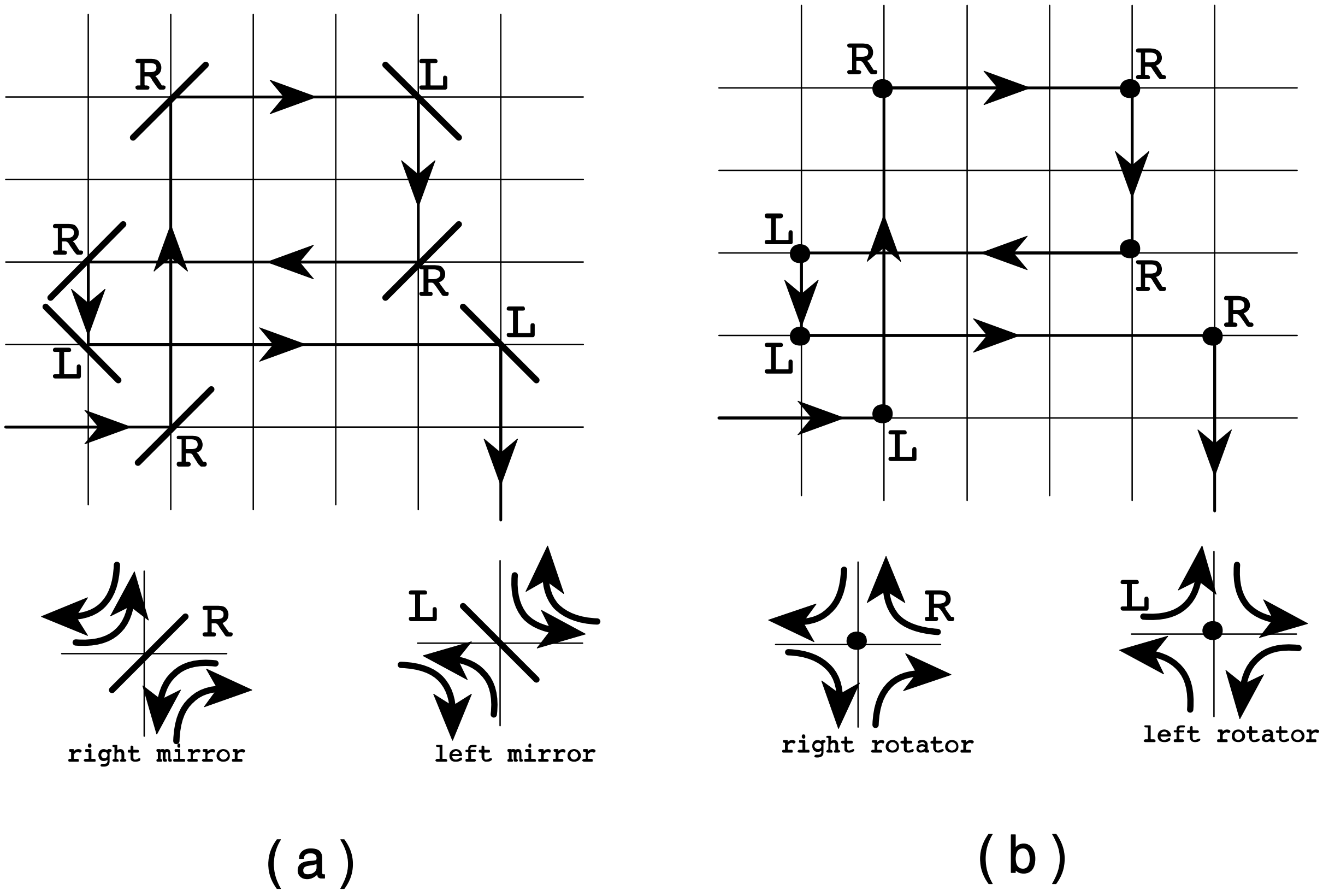

Fig. 1

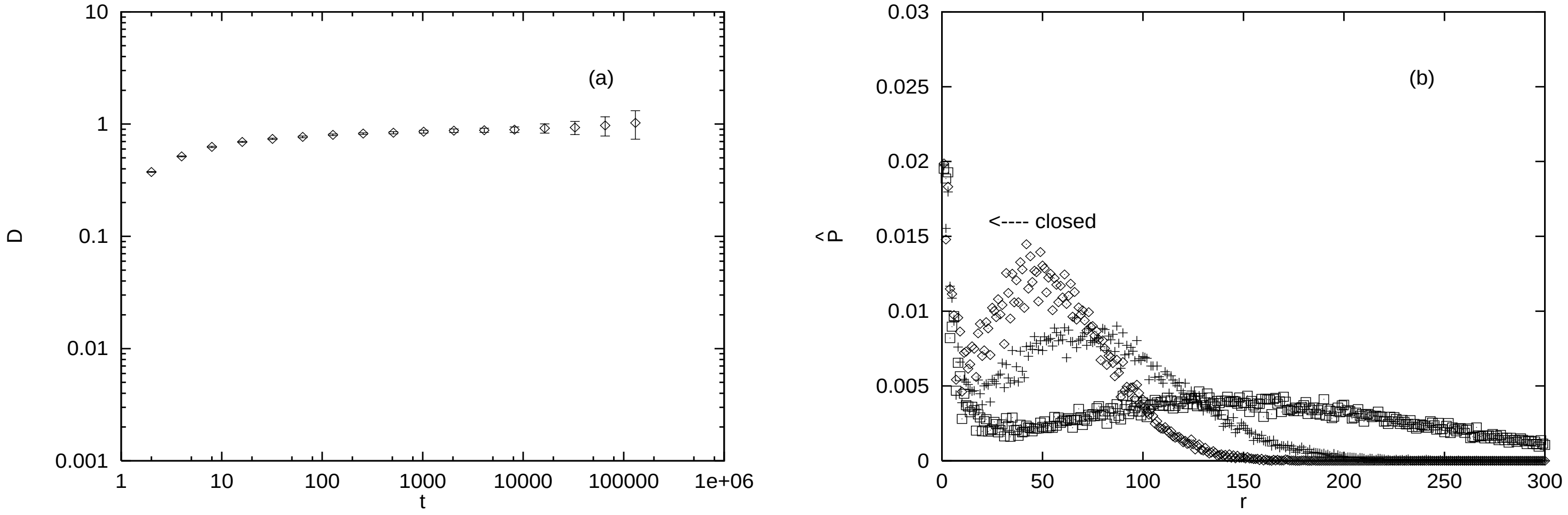

Fig. 2

.

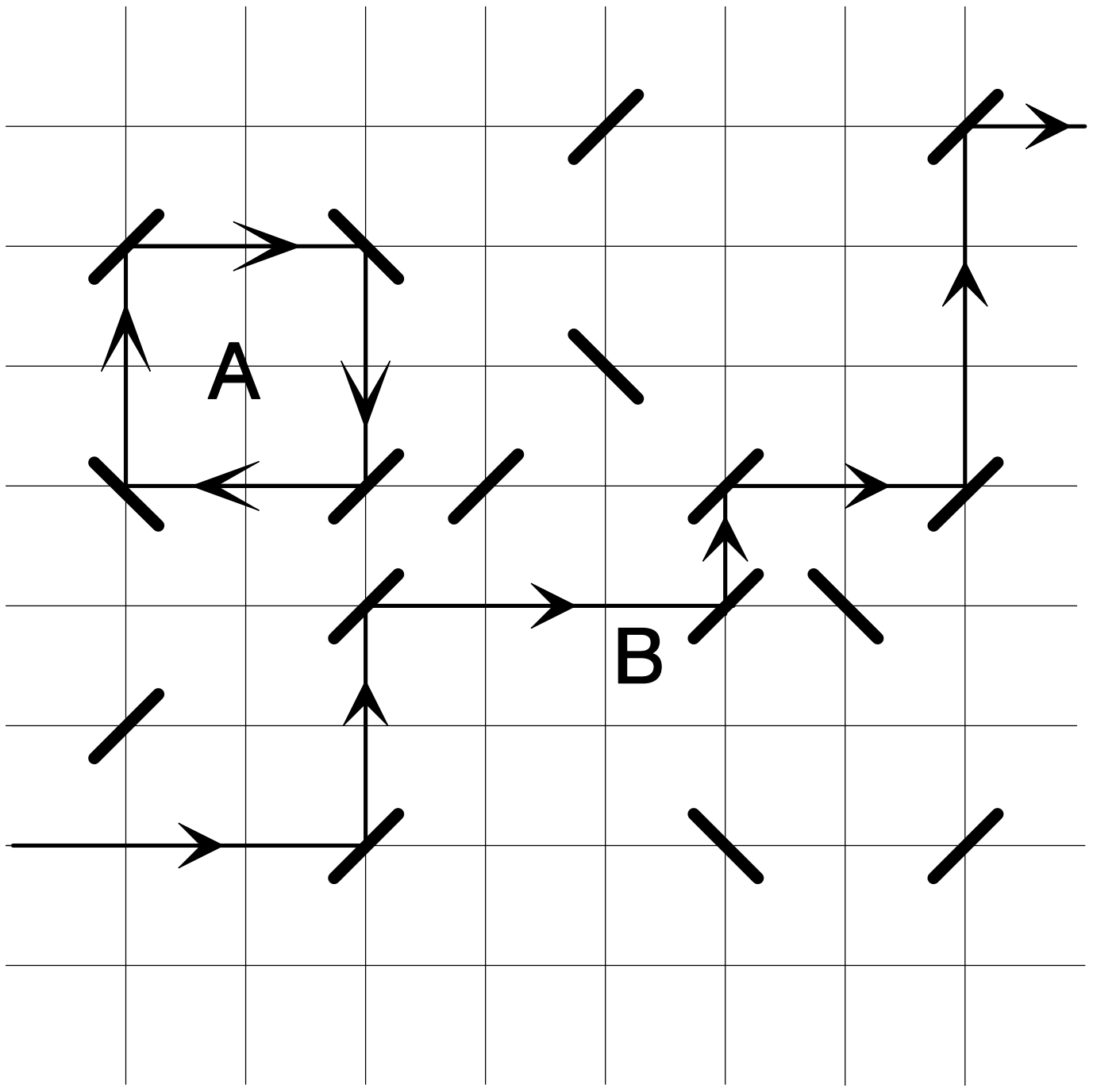

( c )

Fig. 2

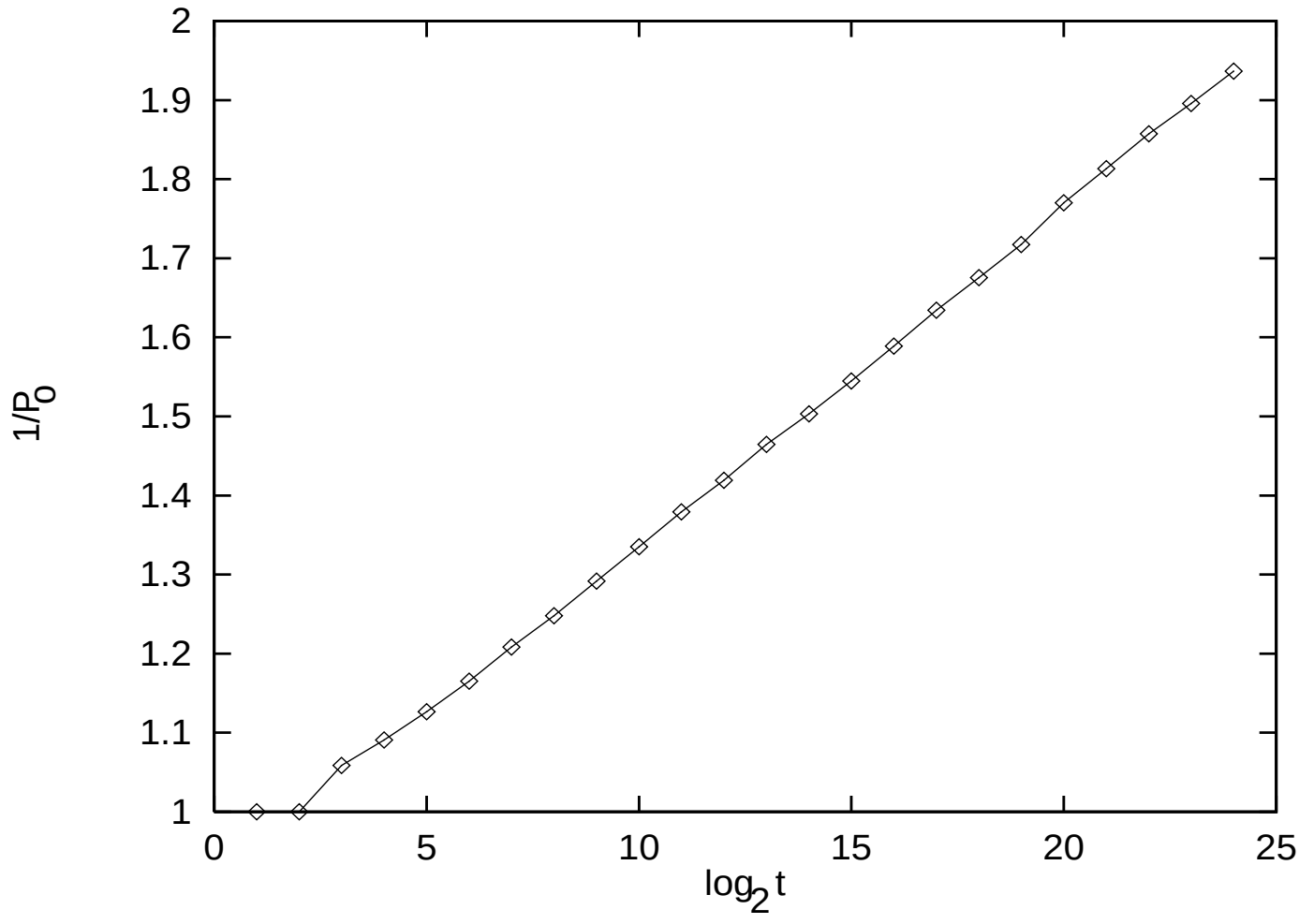

Fig. 3

.

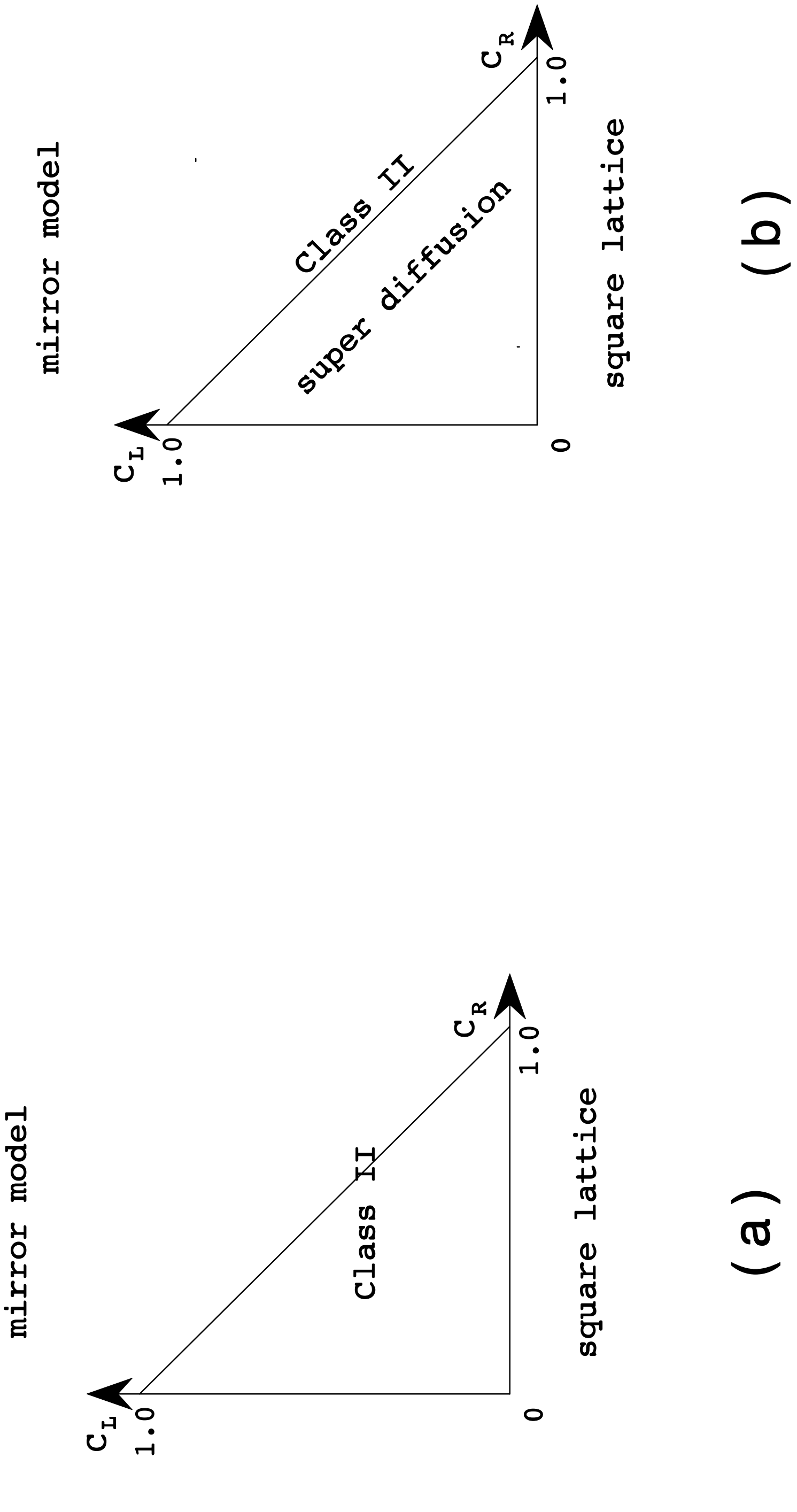

Fig. 4

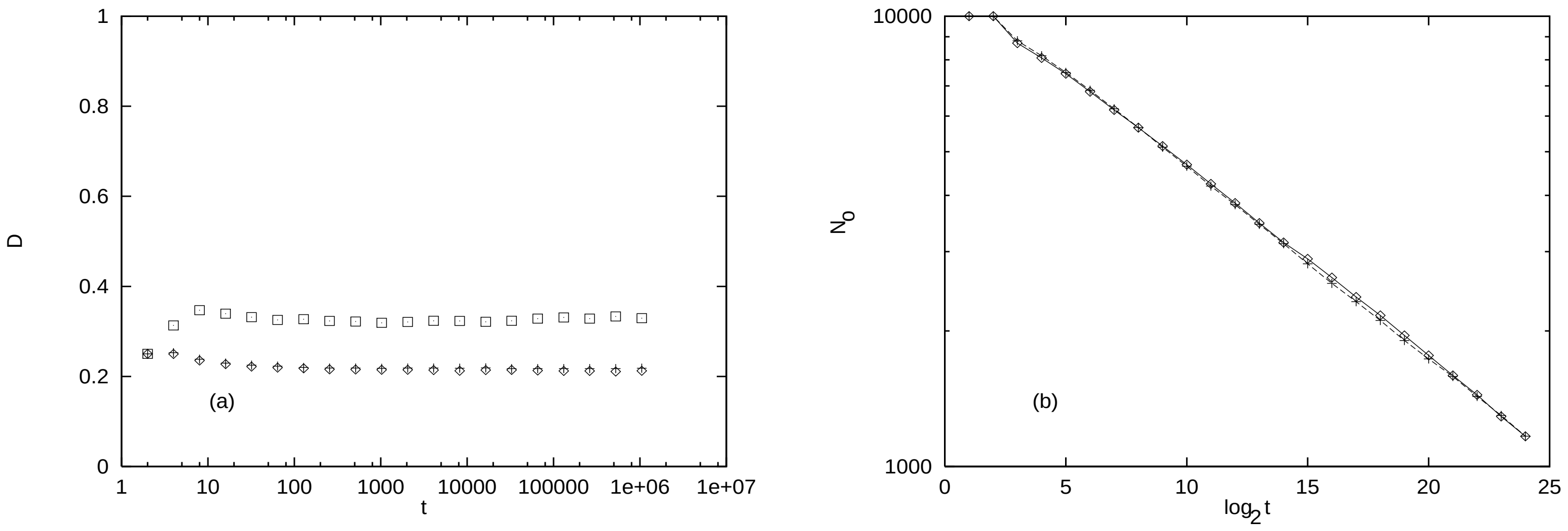

Fig. 5

.

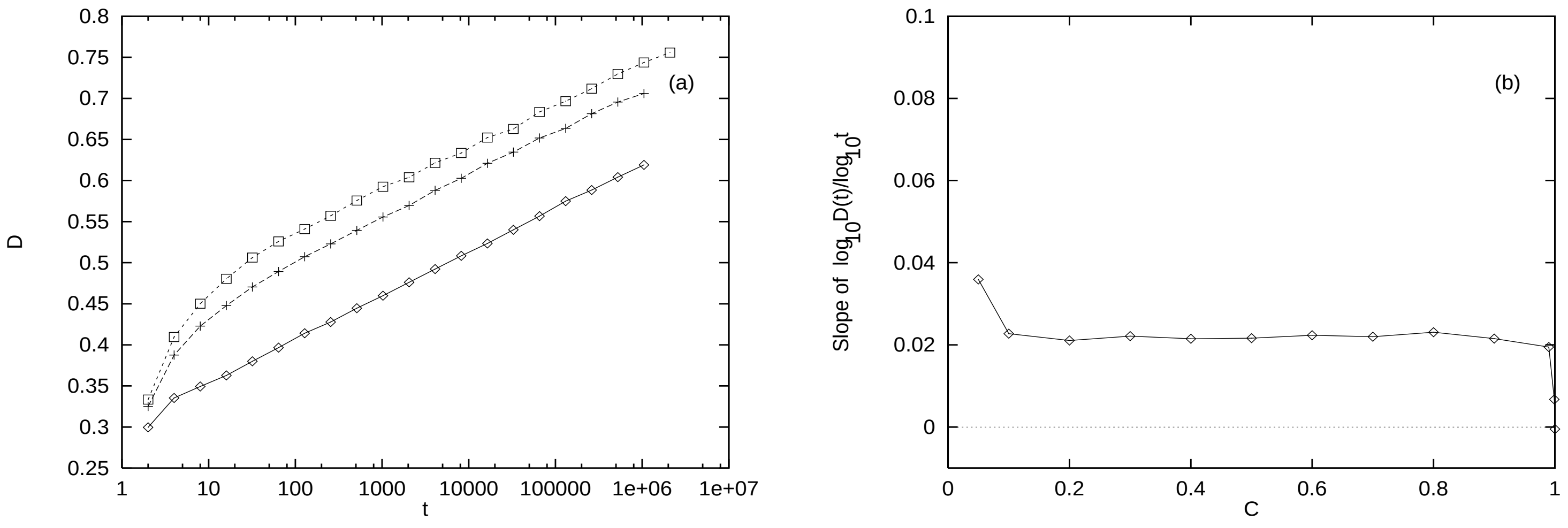

Fig. 6

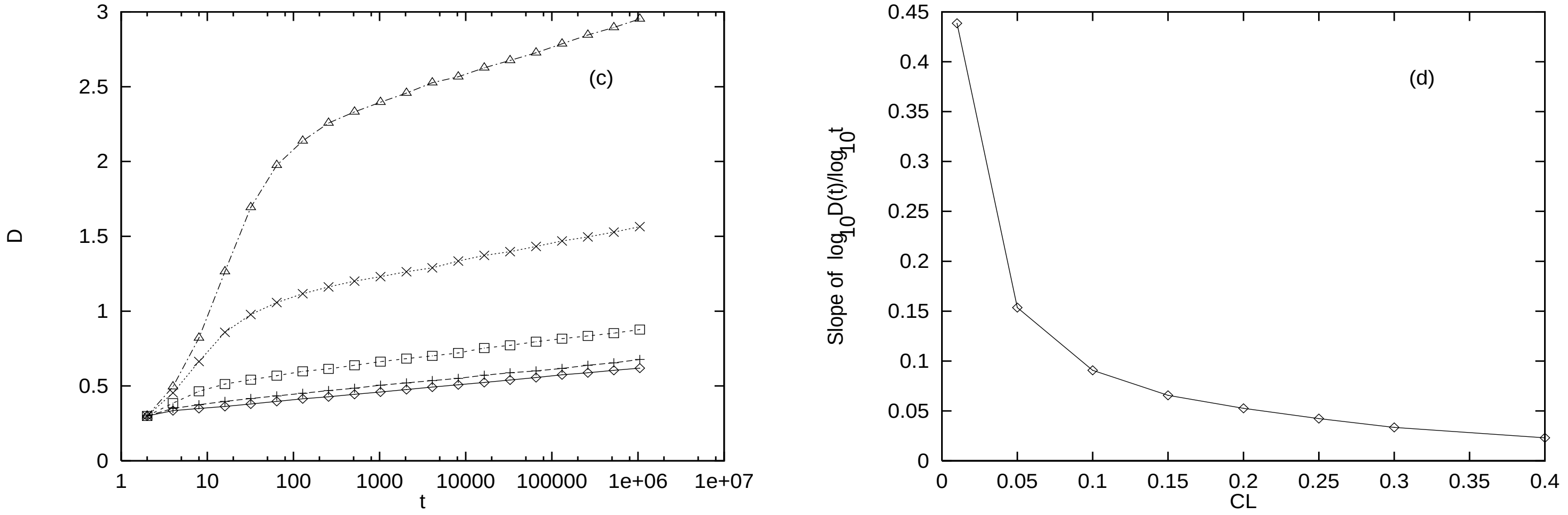

Fig. 6

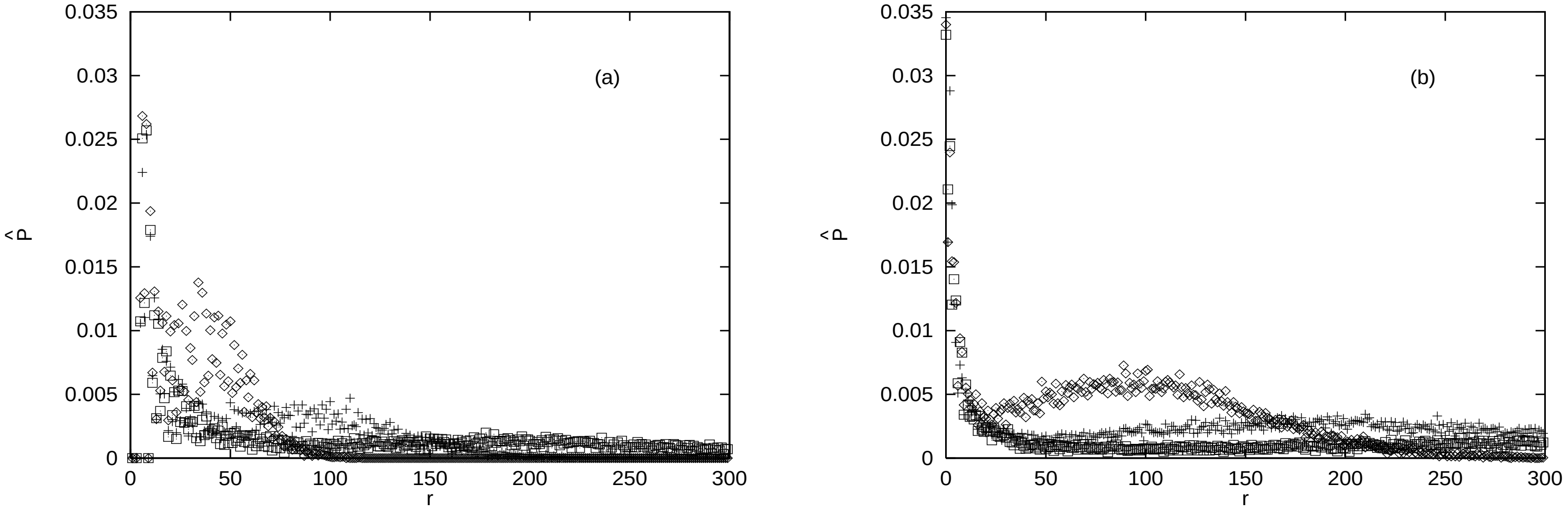

Fig. 7

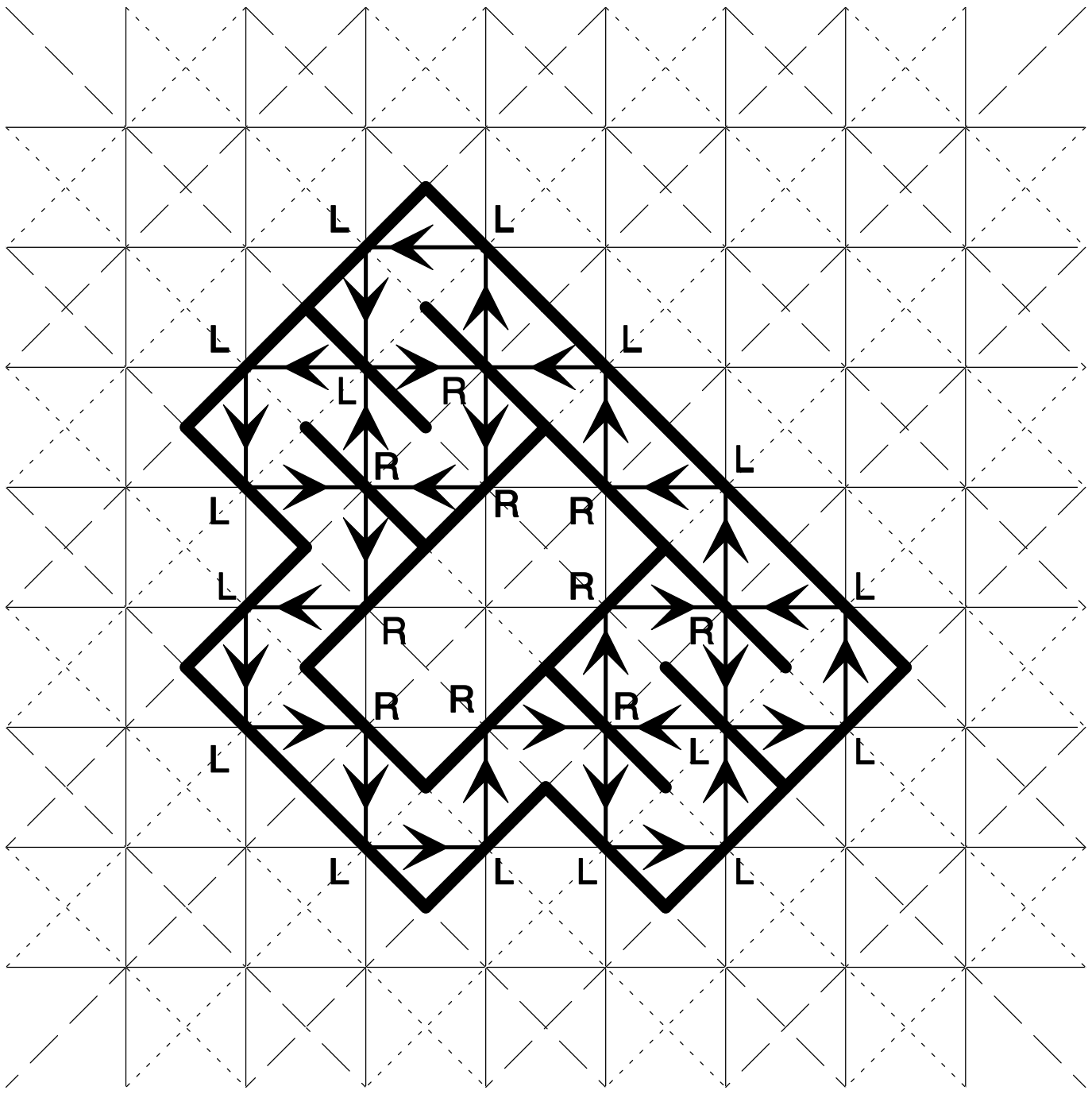

Fig. 8

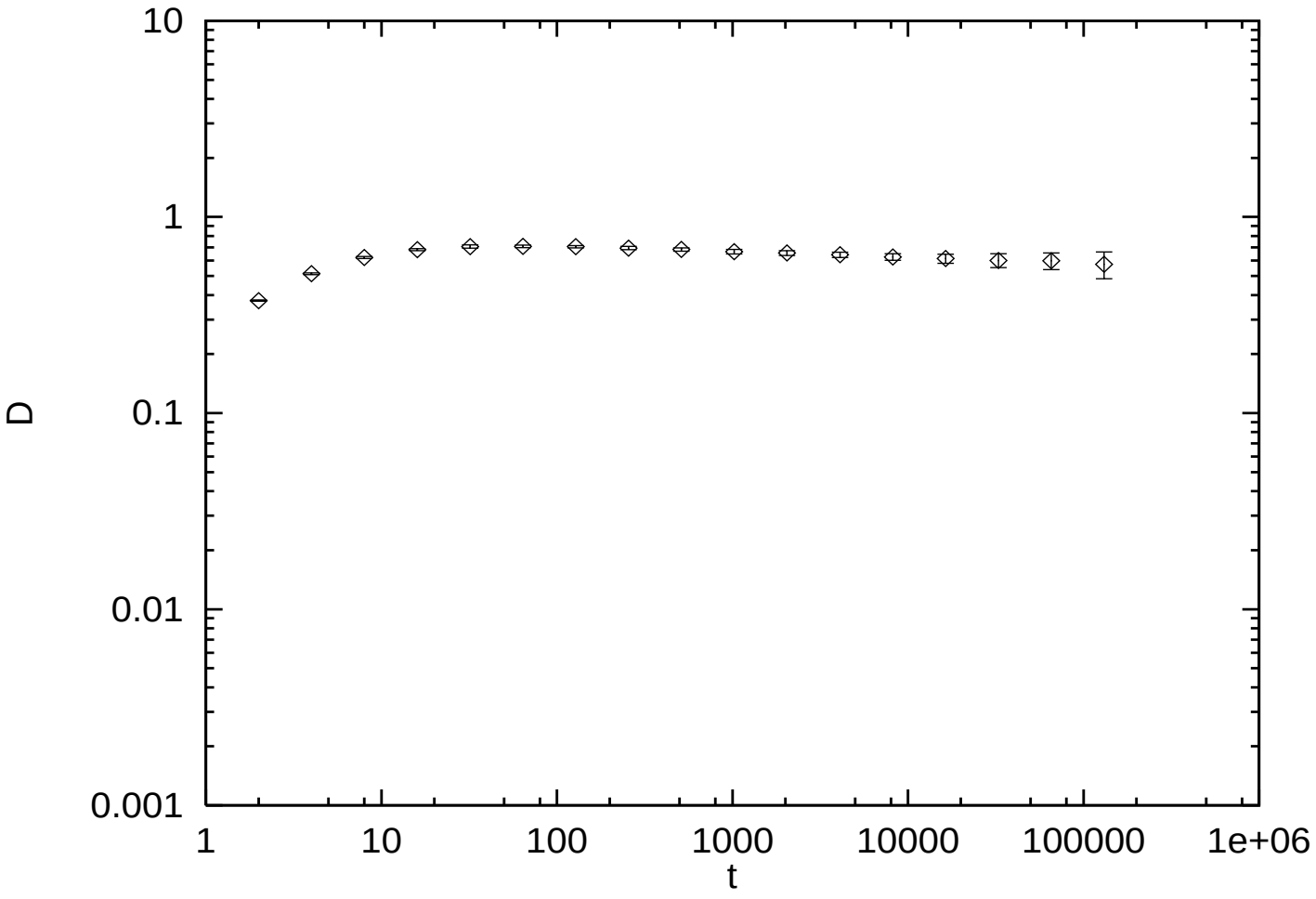

Fig. 9

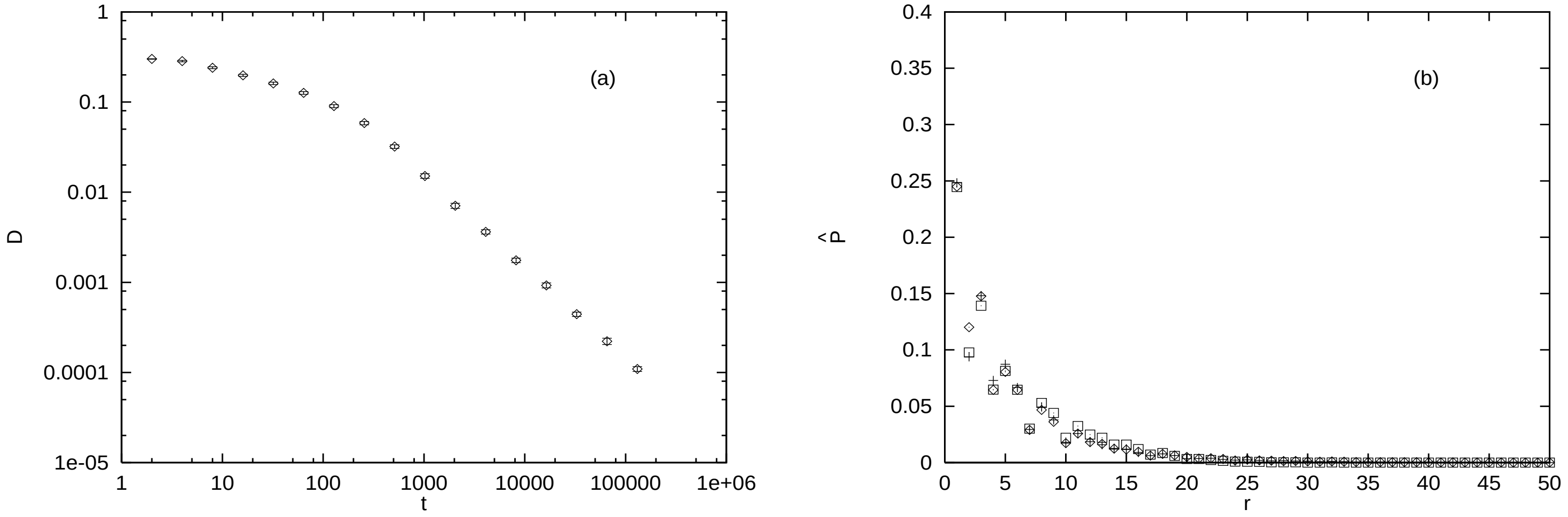

Fig. 10

.

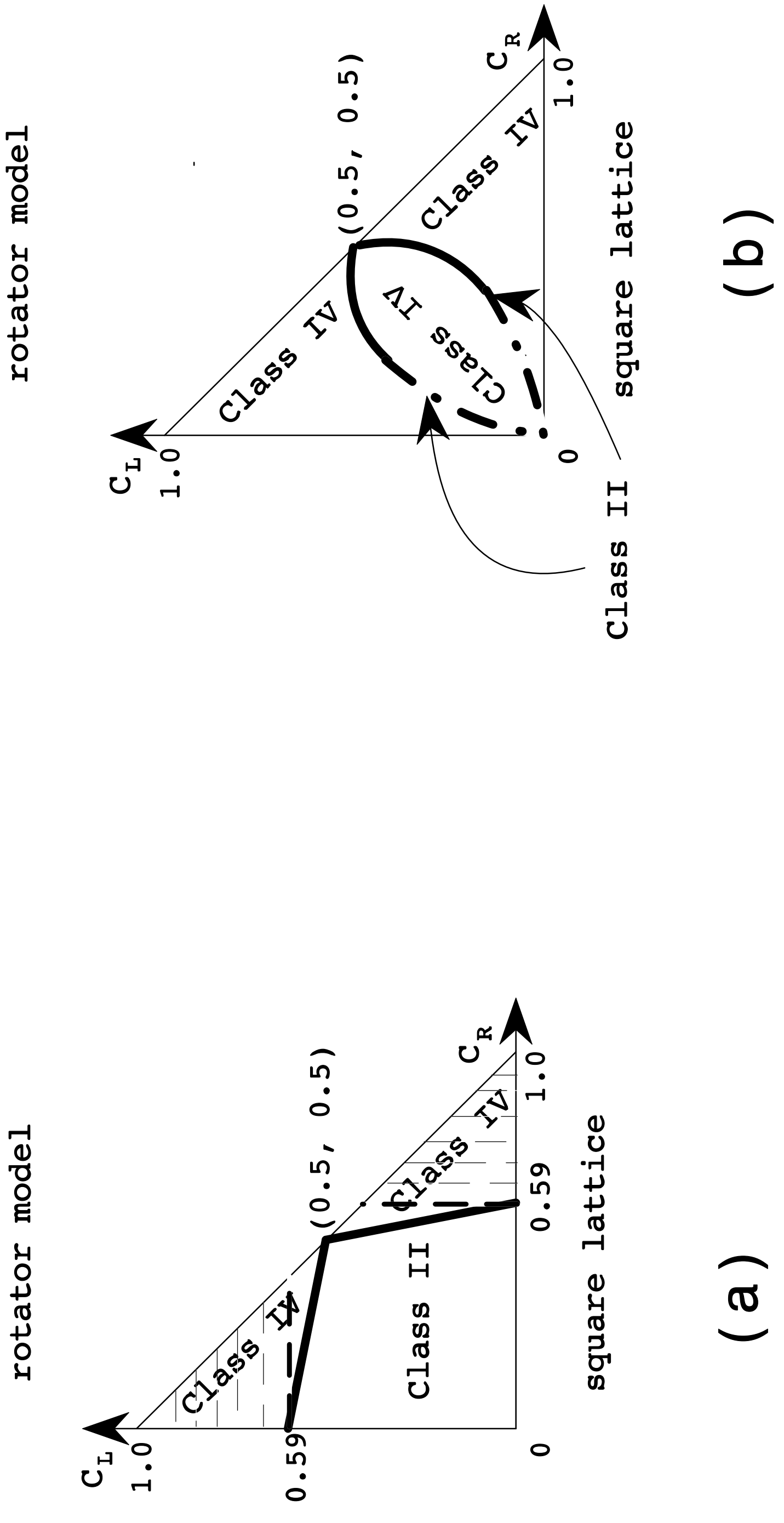

Fig. 11

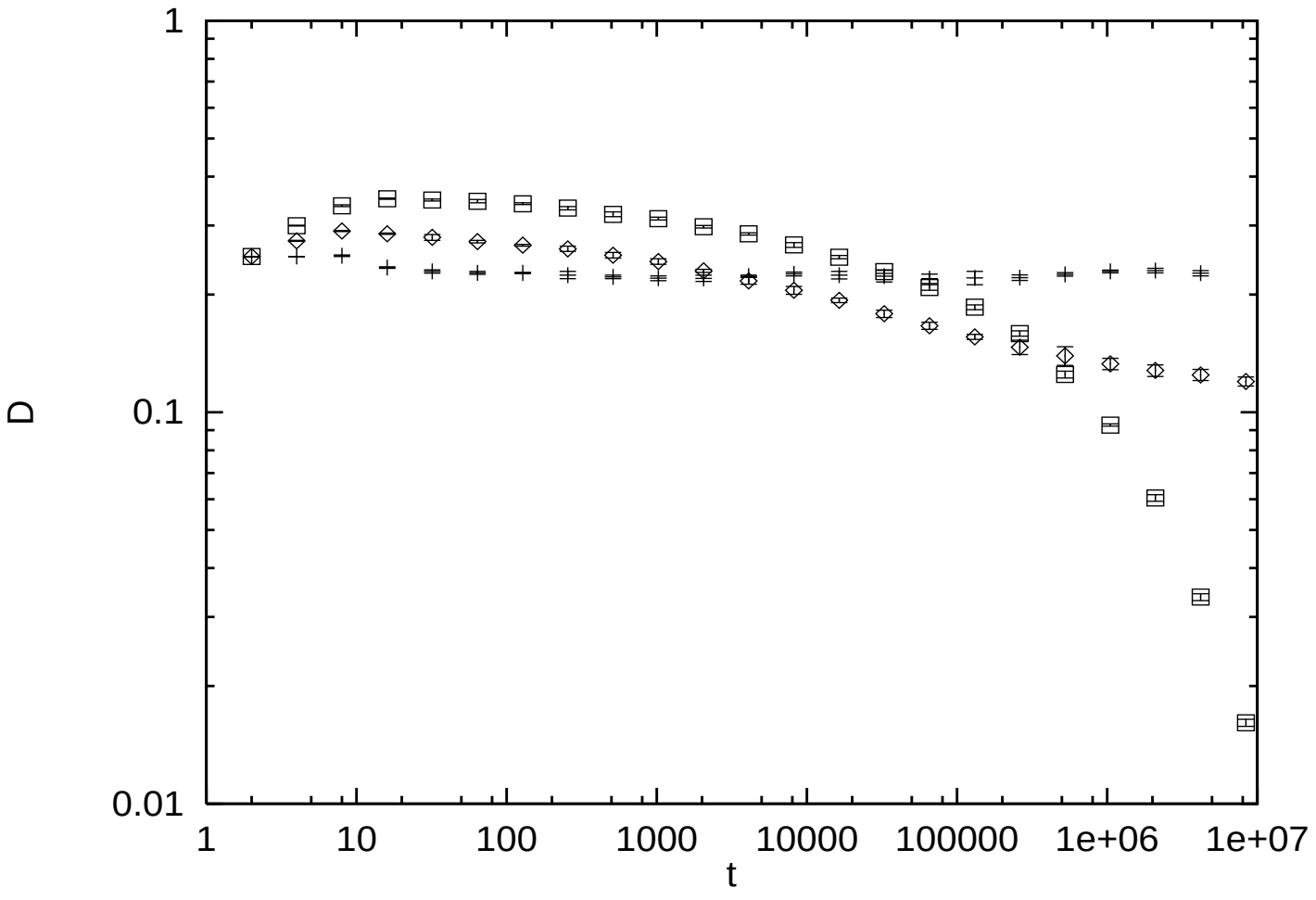

Fig. 12

.

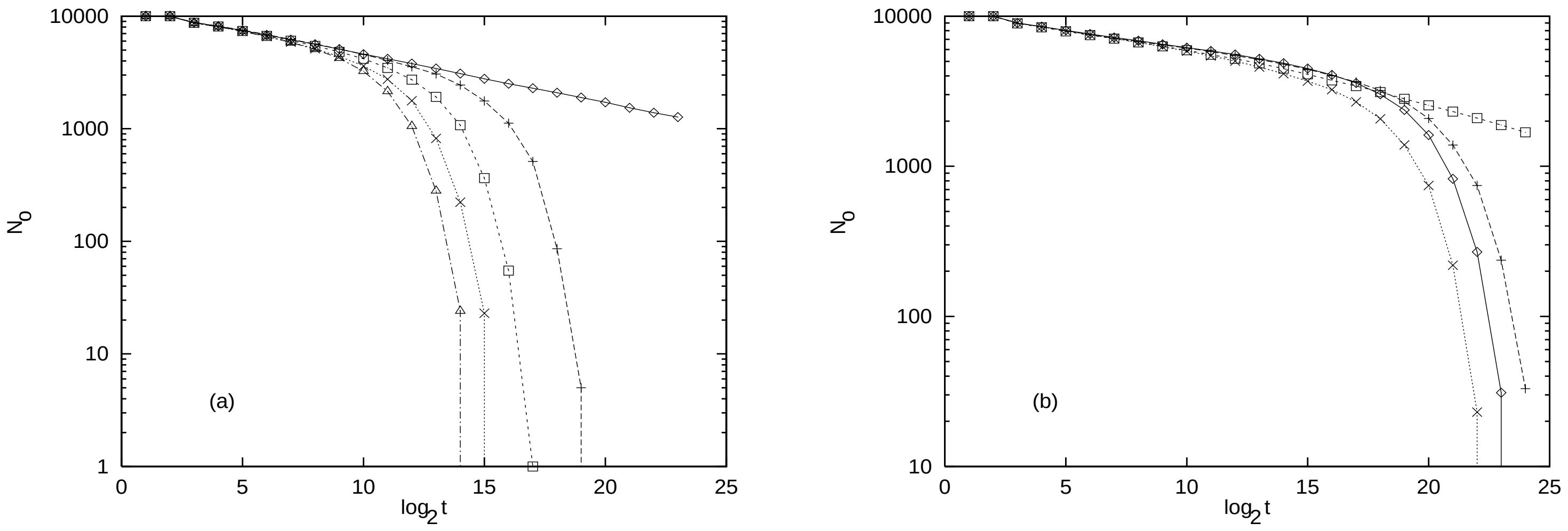

Fig. 13

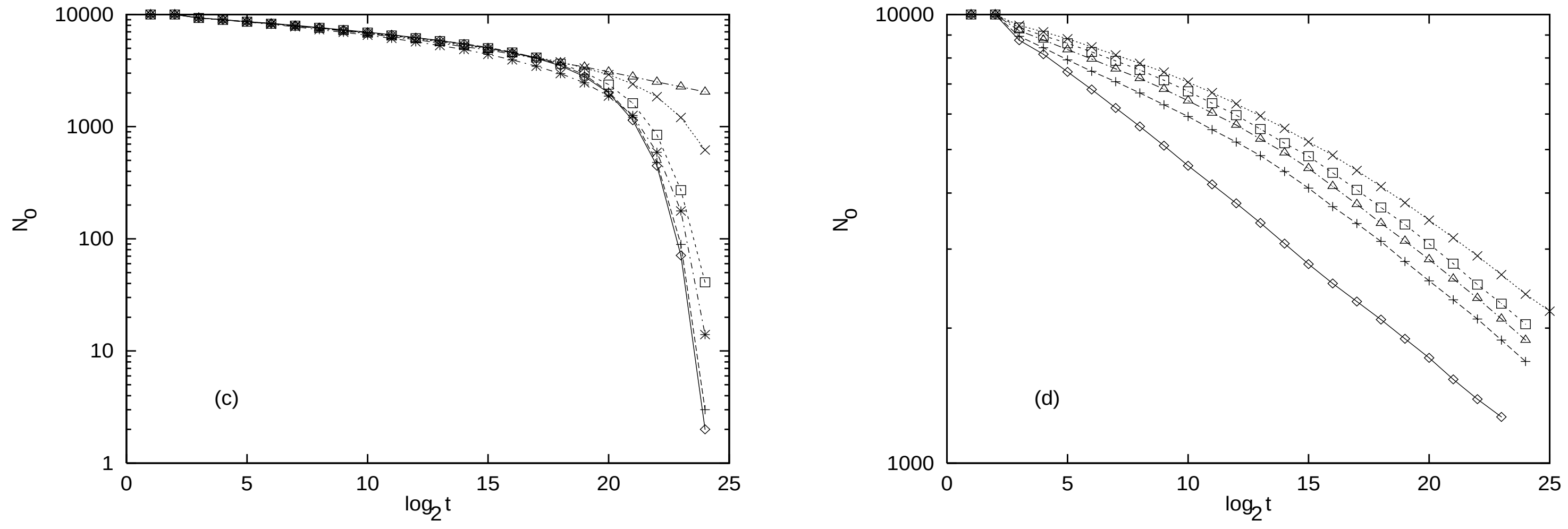

Fig. 13

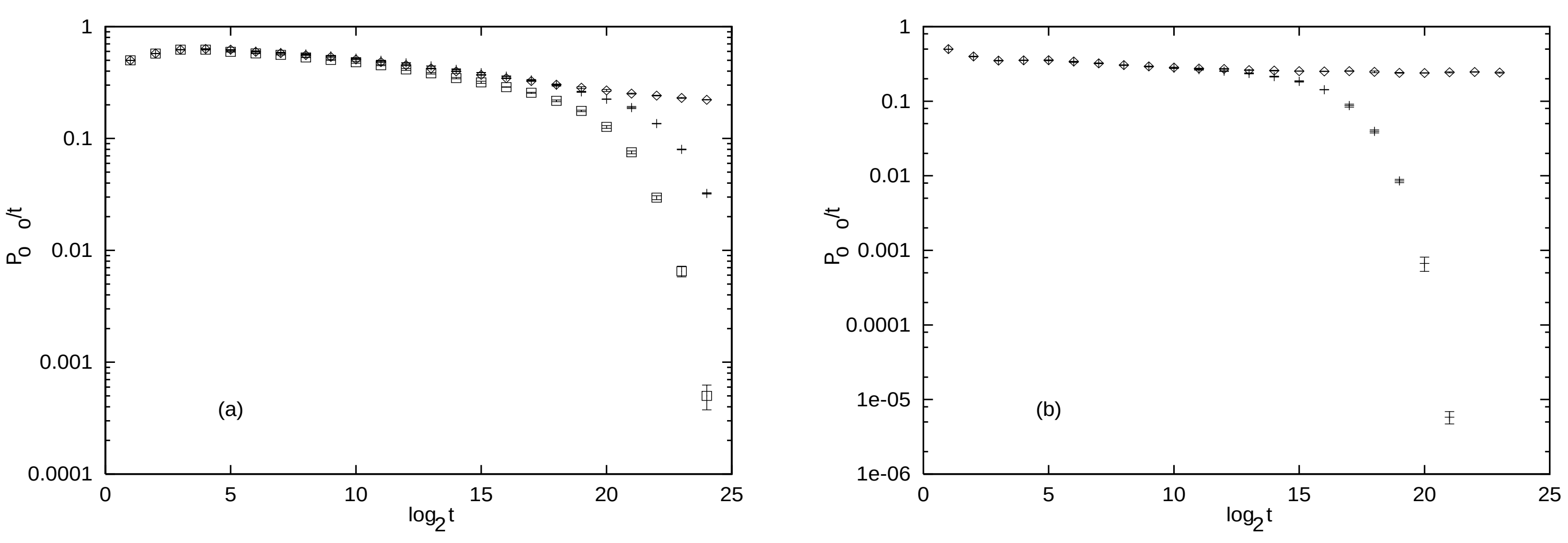

Fig. 14

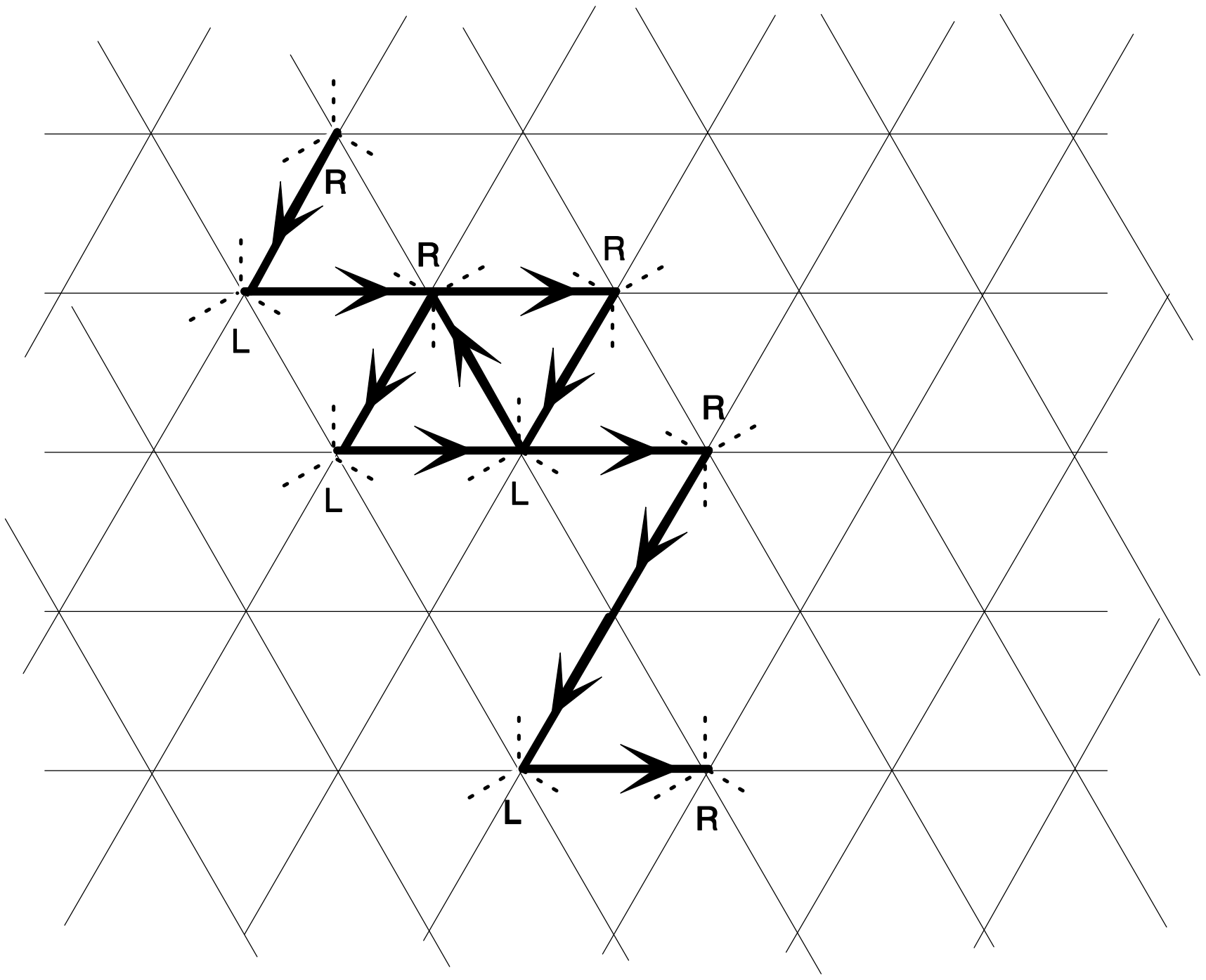

(a)

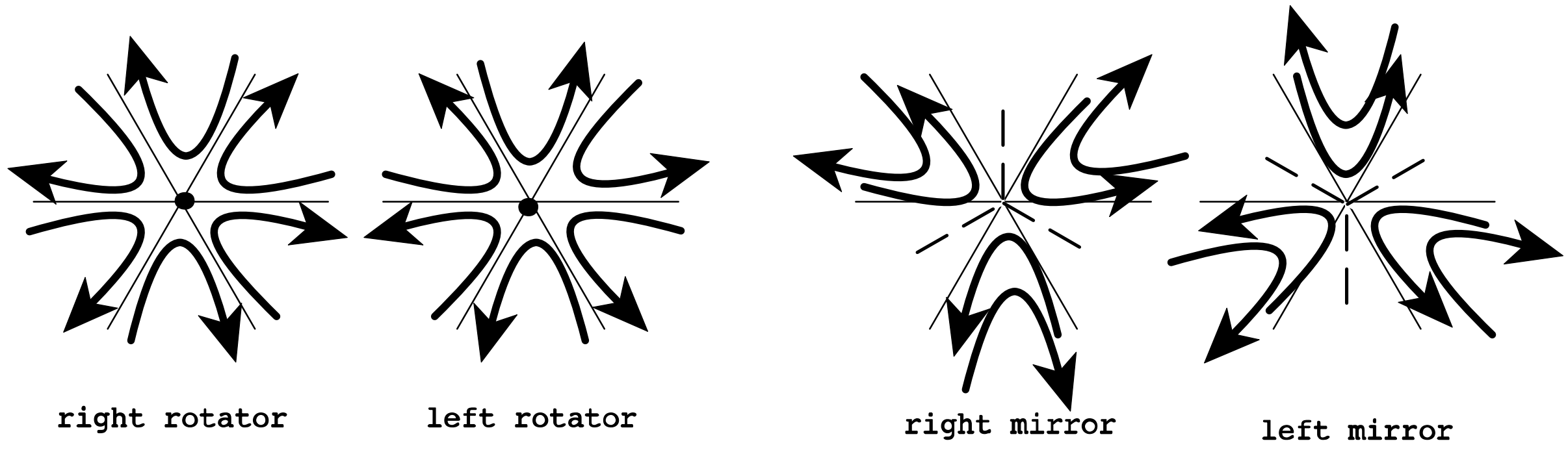
right rotator
left rotator
right mirror
left mirror

(b)

Fig. 15

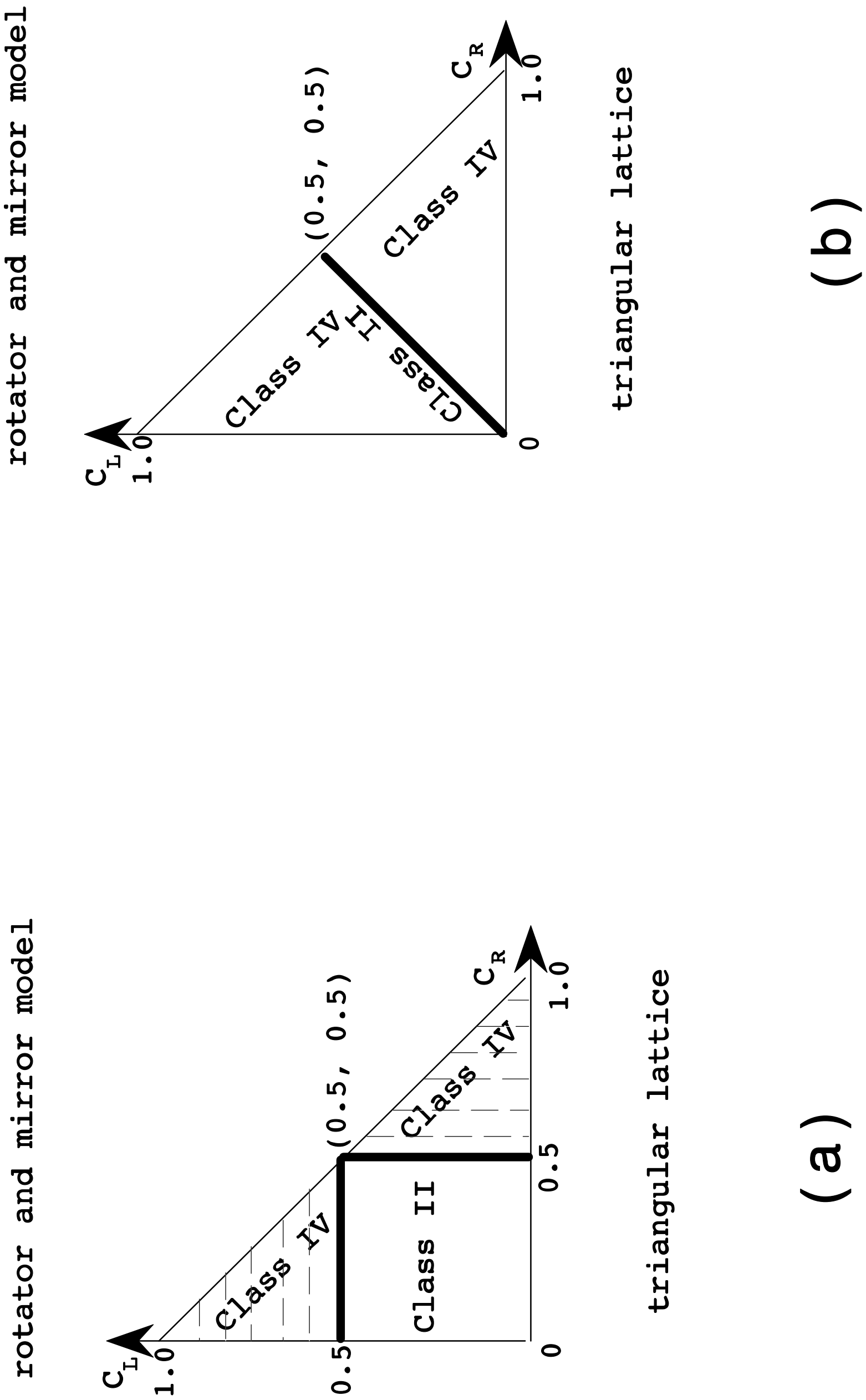

Fig. 16

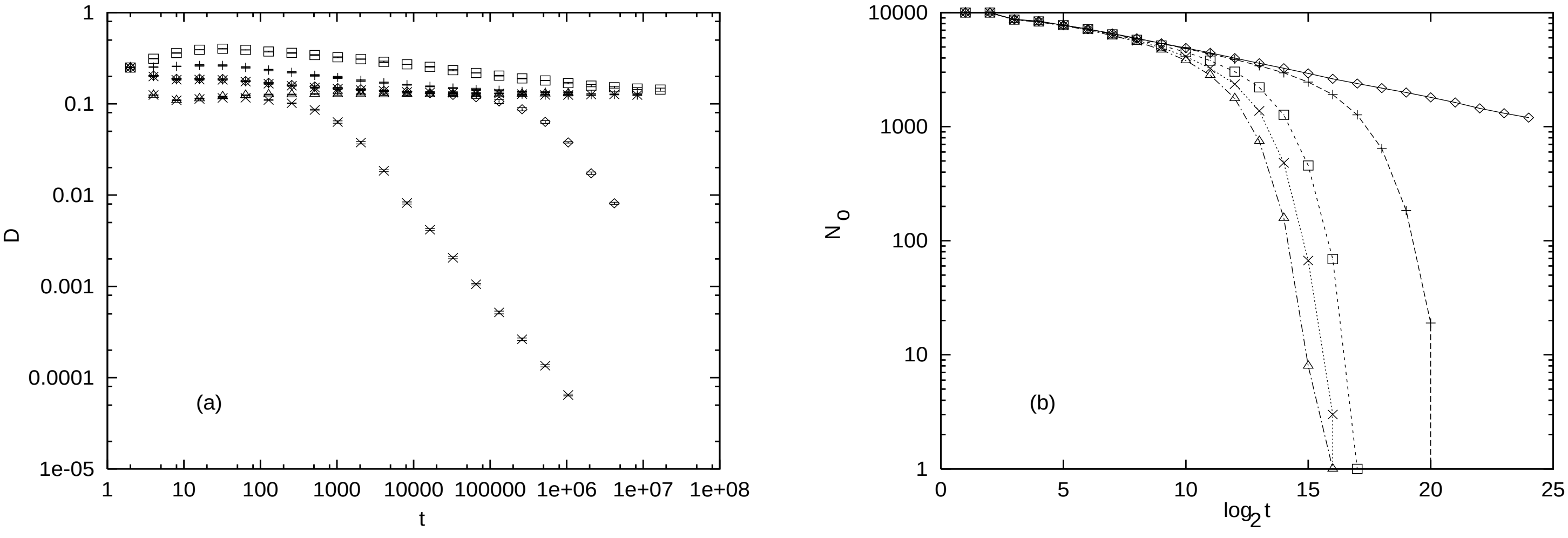

Fig. 17

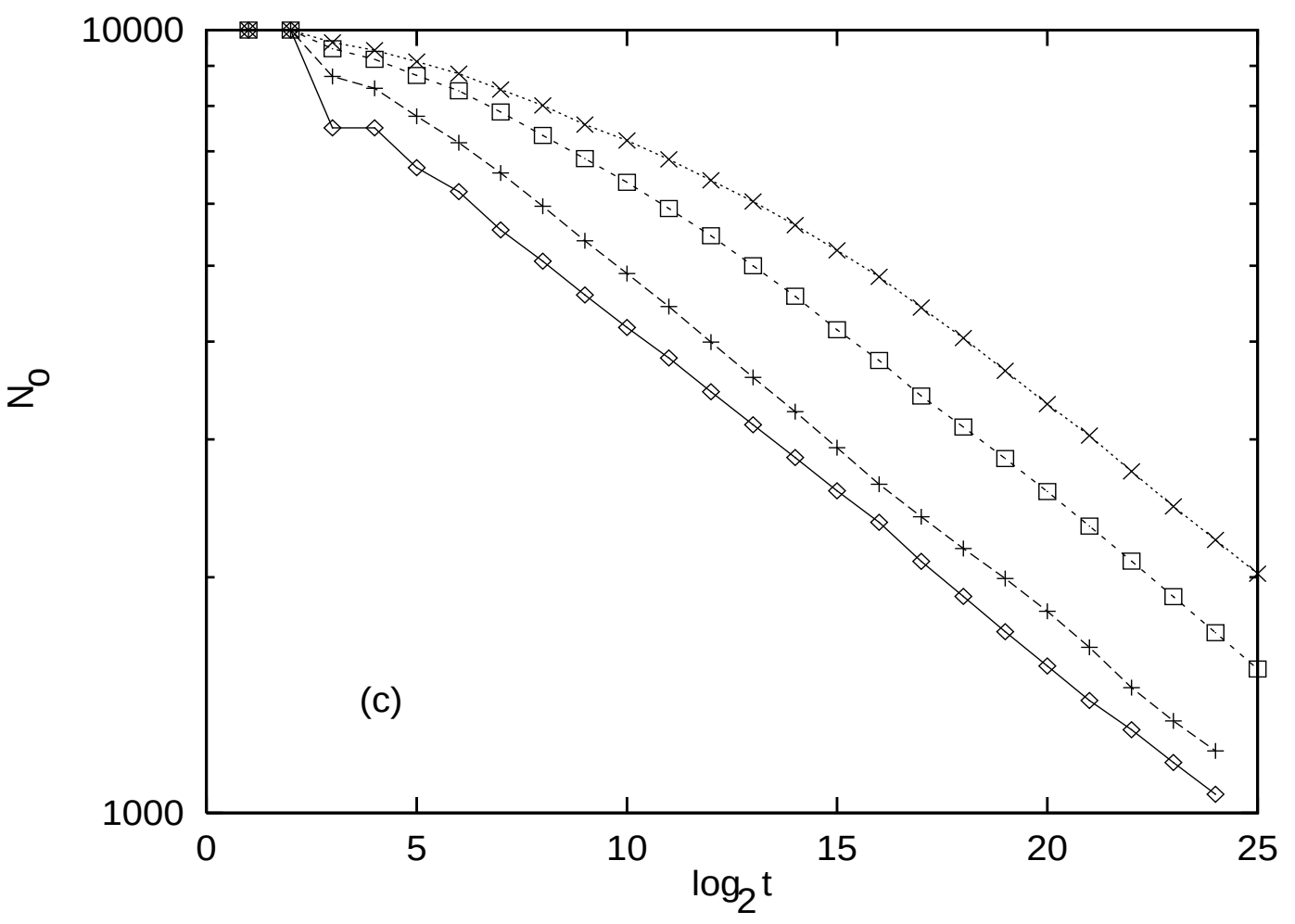

Fig. 17

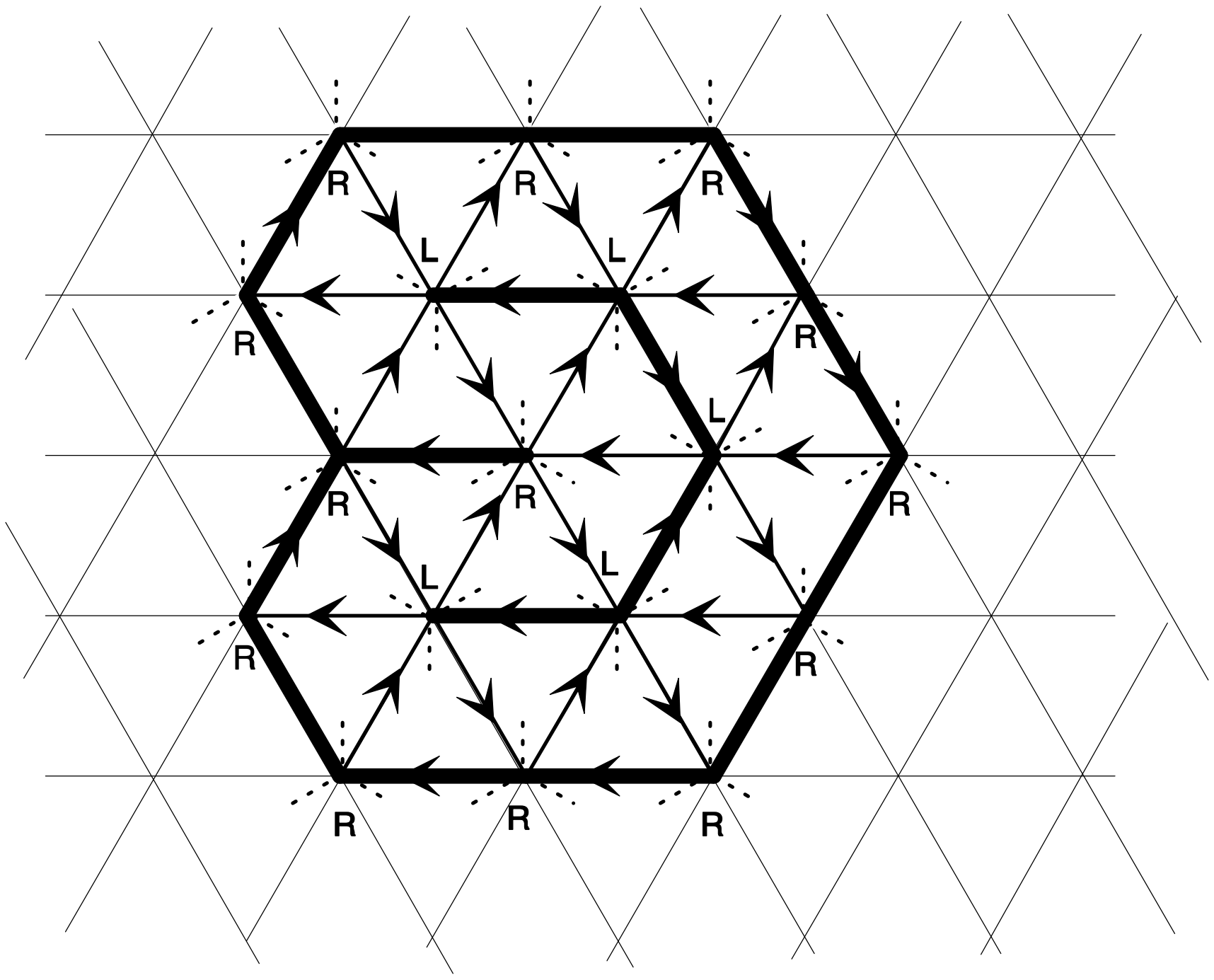

Fig. 18